\documentclass[twocolumn,
superscriptaddress,
 amsmath,amssymb,
 aps,
]{revtex4-2}

\usepackage{todonotes}
\usepackage{comment}
\usepackage{bm}
\usepackage{soul}
\usepackage{color}
\usepackage{tabularx}
\newcommand{\ri}{\mathrm{i}}
\newcommand{\diag}{\mathrm{diag}}

\begin{document}

\title{Learning the Optimal Hydrodynamic Closure}


\author{Florian Kogelbauer}
\email{floriank@ethz.ch}\thanks{Corresponding Author}
\affiliation{Department of Mechanical and Process Engineering, ETH Z{u}rich, 8092 Z{u}rich, Switzerland}

\author{Candi Zheng}
\email{czhengac@connect.ust.hk}
\thanks{C.Z. and F.K contributed equally to this work}
\affiliation{Department of Mathematics, Hong Kong University of Science and Technology, Clear Water Bay, Hong Kong SAR 999077, China}

\author{Ilya Karlin}
\email{ikarlin@ethz.ch}
\affiliation{Department of Mechanical and Process Engineering, ETH Z{u}rich, 8092 Z{u}rich, Switzerland}

 


\begin{abstract}
We present the optimal hydrodynamic model for rarefied gas flows relative to a given kinetic model by combining the recent theory of slow spectral closure with machine learning techniques. We learn generalized transport coefficients from density fluctuation data for the Shakhov model as well as Monte Carlo Simulations and demonstrate that our approach decisively outperforms previously proposed constitutive laws for higher-order hydrodynamics. The novel hydrodynamic model is in close alignment with the underlying kinetic models, thus proving the optimality of the slow spectral closure. Our theory is independent on any smallness assumption of the Knudsen number and is formulated solely in terms of macroscopic observables.
\end{abstract}

\date{\today}

\maketitle

Accurate modeling of multiscale phenomena constitutes one of the most intriguing challenges of modern physics. Systems without a pronounced separation of scales appear in turbulent flows 
\cite{zakharov2012kolmogorov}, chemistry \cite{karplus2014development} and polymer physics \cite{zeng2008multiscale}. One particularly rich and complex family of scale-free dynamics are rarefied gas flows as encountered in the atmosphere \cite{schaaf1961flow} or in  micro-fluidics \cite{akhlaghi2023comprehensive}. Up to now, there is no universal method to obtain general constitute laws for fluid dynamics at any rarefaction level from molecular models that would be reliable yet easy to calculate. In this work, we demonstrate that, despite a lack of an immediate scale separation in kinetic theory, the governing equations allow for a hidden scale separation which only becomes apparent through a detailed spectral analysis - independent of any smallness assumption on the Knudsen number. This allows us to derive and subsequently learn the dynamically optimal constitutive laws for rarefied hydrodynamics on the linear level.

It is well-known that simulations of rarefied gases are notorious for non-negligible influences of the mesoscopic scale on the overall dynamics. Due to difficulties in solving the full Boltzmann equation directly, computations for rarefied gases heavily rely on statistical methods such as Direct Simulation Monte Carlo (DSMC) \cite{Bird1994MolecularGDF} or  Fokker--Planck methods \cite{jenny2010solution}, which are costly and time-intensive. Even simplified kinetic models, such as the Shakhov model \cite{shakhov1968generalization}, are expensive to simulate. To tackle these difficulties, variants of extended hydrodynamics are typically coarse-grained or projected onto finitely many moments
to make computations more feasible, which, however, limits their applicability to weak rarefaction regimes \cite{struchtrup2003regularization,wu2020accuracy}. A different approach to obtain extended extended hydrodynamicist relies on the Mori--Zwanzig formalism and involves temporal integral contributions \cite{zwanzig2001nonequilibrium}. 

Apart from computational efficiency, the fundamental physical question of consistent models across scales, i.e., the connection between kinetic theory and extended hydrodynamics, lingers on for over a century and further emphasizes the need to bridge our understanding from the microscopic Newtonian description of matter to the macroscopic formulation of continuum mechanics as famously insinuated by Hilbert \cite{hilbert2000mathematical}. The convergence of solutions of the Boltzmann equation to solutions of the Navier--Stokes equation for vanishing Knudsen number marks a milestone in our understanding of Hilbert's sixth problem \cite{saint2009hydrodynamic} and gives the correct constitutive laws for the fully fluidic regime. Despite intensive research, however, it remains a superb challenge to obtain physically sound, yet low-dimensional macroscopic dynamics from a given microscopic model for all levels of rarefaction \cite{wu2020accuracy}.

Classically, higher-order hydrodynamics are derived from kinetic models through the Chapman--Enskog series \cite{chapman1990mathematical} - a Taylor expansion in Knudsen number - and Grad-type systems by projection onto moments \cite{grad1949kinetic}. Higher-order hydrodynamics, such as the Burnett equation \cite{burnett1935distribution}, however, develop nonphysical instabilities \cite{bobylev2006instabilities}. While the lack in hyperbolicity may be remedied via Bobylev regularization  \cite{bobylev1982chapman}, Chapman--Enskog- and Grad-type approximations are inherently limited to small Knudsen numbers and do not offer any insight in the definition of hydrodynamic entropy from first principles \cite{SLEMROD20131497}. More dramatically, the Chapman--Enskog series might even be divergent altogether \cite{kogelbauer2025relation}. Conceptually, the deficiencies of the Chapman--Enskog series results from applying a formal Taylor series for a singularly perturbed system, while the deficiencies of the Grad projection result from a finite-dimensional truncation of an infinite-dimensional operator in moment space. Both assume a certain proximity to global equilibrium that prevent their extension to higher rarefaction levels. How can we overcome these deficiencies?

The key idea of this work is to apply the theory of slow manifolds to kinetic equations on the spectral level in combination with machine learning techniques to give a dynamically optimal solution to the moment closure problem. Historically, the search for invariant manifolds in kinetic dynamics dates at least back to Hilbert himself, while recently, solutions to the invariance were investigated analytically and numerically \cite{gorban2005invariant, gorban2014hilbert}.\\
The theory of slow manifolds, which originated from atmospheric sciences \cite{lorenz1992slow}, seeks a distinguished invariant manifold in phase space which attracts all trajectories in a neighborhood exponentially fast \cite{mackay2004slow}. Slow manifolds are widely applied in model reduction \cite{roussel2001invariant,axaas2023fast} and have classically been associated with geometric singular perturbation theory \cite{fenichel1979geometric}, relying on a small parameter that defines the slow dynamics. Rather recently, the existence of slow manifolds was shown rigorously with the parametrization method \cite{de1997invariant,cabre2003parameterization}, relying on spectral information of the linearization around a global equilibrium. Indeed, slow manifolds derived through eigenvalues with maximal negative real part do not rely on the expansion in a small parameter and can thus capture dynamics for which series expansions and moment projections fail.




The current work is built upon the construction of the slow manifold based on information of the operator spectrum \cite{ellis1975first}, which allows us to derive explicit formulas for the generalized transport coefficients in terms of spectral quantities \cite{PhysRevE.110.055105}, much rather than numerical solutions for the generalized transport coefficients. The dynamically optimal hydrodynamic closure thus combines the best of both worlds: It is formulated in terms of macroscopic fields, which allows for a full description of the fluid with few degrees of freedom, while it is able to capture rarefaction effects 
\cite{PhysRevE.110.055105}
as accurately as a kinetic model, thus rendering it an ideal method for numerical computations of rarefied gas flows. The spectral closure is thus dynamically optimal in the sense that any other closure assumption will be worse when compared to the moments of actual solutions to the underlying kinetic model.



The construction of the slow manifold based on spectral information, however, revealed analytical limitations to a certain range of wave numbers as a prominent feature of exact hydrodynamics, called \textit{criticality}. Furthermore, the detailed and explicit spectral analysis of a linear kinetic operator is often out of scope. In this work, we apply machine learning to overcome the potential limitations of criticality, thus allowing for the derivation of constitutive laws over an unprecedentedly large range of Knudsen numbers.  

\begin{figure}[h!]
    \centering
\includegraphics[width=0.85\linewidth]{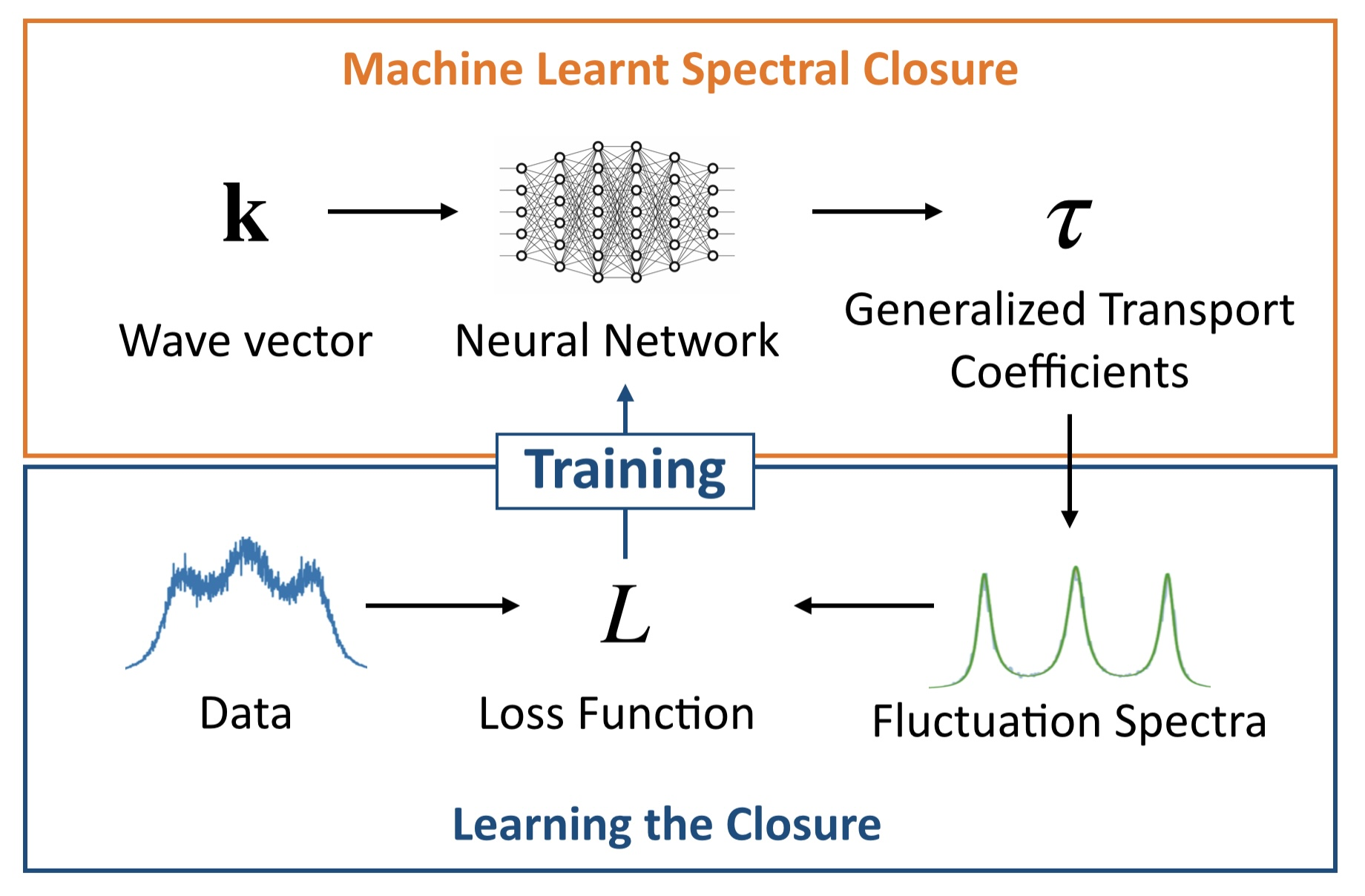}
    \caption{Schematics of the learning algorithm: The neural network is trained on density fluctuation data. A wave vector as an input is mapped to  the generalized transport coefficients as an output, which in turn define the predicted density fluctuation spectra.
}
    \label{neuralnet}
\end{figure}

Recently, the use of modern machine learning techniques, especially neural networks, opened up 
a promising direction to derive constitutive laws from data \cite{zhou2021learning}.  
While prior approaches were focused on learning dynamical models for physical processes \cite{Hana2019}, we aim for a framework to learn constitutive equations, i.e., the optimal dependence of higher-order moments on lower-order moments \cite{zheng2023data} directly from density-fluctuation data, see Figure \ref{neuralnet}. In particular, we contrast the learning of constitutive laws to Physics Informed Neural Networks (PINNs) to solve the Boltzmann equation \cite{lou2021physics}, which learn a partial differential equation by minimizing cost functions on trajectory data, assuming a certain, finite set of dictionary functions \cite{raissi2019physics}. As PINNs do not learn a constitutive law, they cannot connect kinetic theory to fluid dynamics and thus cannot tackle Hilbert's sixth problem.

We also delineate our work from neural operators, which learn the governing equations of a physical process as a neural network and try to solve it subsequently by time integration. This learning from realized trajectory data is inherently limited to small time scales and lacks long term evolution stability \cite{Hana2019}. Both PINNs and neural operators struggle with explicitly identifying the dependence of solutions on Knudsen number. 

In this work, we learn the dynamically optimal constitutive laws as predicted by the theory of spectral closure \cite{PhysRevE.110.055105}. We use machine learning to overcome the analytical restriction of criticality and the explicit calculation of eigenvalues in the theory of slow manifolds of the Boltzmann equation, see Section \ref{LearningHydroSec} for details. Our approach naturally allows for the long-time evolution of initial conditions. while preserving the flexibility of learning. Figure \ref{neuralnet} gives an overview of our approach to learning dynamically optimal constitutive laws. 





The paper is structured as follows. In Section \ref{MomentClosureProblem}, we recall the moment closure problem for kinetic equations, while in Section \ref{SpectralClosureSec} we summarize the properties of the optimal hydrodynamics as derived from the spectral closure. Section \ref{LearningHydroSec} explains how the optimal hydrodynamics can be learned from density fluctuation data. Finally, in Section \ref{Results}, we  use the structural properties of the transport matrix in frequency space to define a low-parametric learning scheme for the generalized transport coefficients. We demonstrate the optimality of the spectrally-closed hydrodynamics by comparing it to the full kinetic model and to three-dimensional DSMC simulations, as well as to the Navier--Stokes equation and the R13 model. The main results of this work are shown in Figures \ref{FigComparison}, \ref{FigComparisonV} and \ref{Figadvection}. We conclude with a general discussion in Section \ref{DiscussionSec}.

\section{The Moment Closure Problem for Kinetic Equations}
\label{MomentClosureProblem}

We consider a general linear kinetic model,
\begin{equation}\label{maineq}
    \frac{\partial f}{\partial t} + \bm{v}\cdot\bm{\nabla} f = \frac{1}{\rm Kn}Q[f],
\end{equation}
for an unknown distribution function $f$,
the linear collision operator $Q$ and the Knudsen number Kn. 
Models of the form \eqref{maineq} typically arise from linearizing a kinetic equation around a global Maxwellian \cite{cercignani1988boltzmann} or as the Fokker--Planck equation from stochastically forced differential equations \cite{risken1996fokker}. As we are interested in fluctuation fields derived from the light-scattering experiment, linear kinetic theory \eqref{maineq} is sufficient to capture the fluctuation spectra and leading-order decay towards equilibrium.

The macroscopic variables density $\rho$, velocity $\bm{u}$ and temperature $T$ are recovered from the distribution function by taking moments according to 
\begin{small}
\begin{equation}\label{macrovariables}
  \rho = \int {m}fd\bm{v},\ \rho\bm{u} =  \int {m} \bm{v} fd\bm{v},\ 
    \frac{3k_BT\rho}{2m}=\int m\frac{(\bm{v}- \bm{u})^2}{2} fd\bm{v},
\end{equation}
\end{small}
where $m$ is the molecular mass and $k_B$ is Boltzmann's constant, and linearizing around a reference state.  
One of the fundamental problems in kinetic theory, the infamous \textit{moment closure problem}, derives from the inability to write the full dynamics in \eqref{maineq} in terms of the macroscopic fields in \eqref{macrovariables} alone. Indeed, due to the transport term $\bm{v}\cdot\bm{\nabla}$, higher-order moments, such as the stress tensor and the heat flux, will enter the moment dynamics inevitably, see Appendix \ref{densityspectra}. This leads to an infinite chain of moment equations, which needs to be closed by a constitutive law for the higher-order fluxes. As detailed in \cite{PhysRevE.110.055105}, the slow manifold assumption leads to a dynamically optimal solution to the closure problem, which we will detail in the following section. As mentioned before, dynamically optimal means that any other closure assumption will deviate more from a general solution of the underlying kinetic model.


\section{Spectral Closure and Generalized Transport Coefficients}
\label{SpectralClosureSec}



\begin{figure*}
    \centering
    \includegraphics[width= 17.8cm]{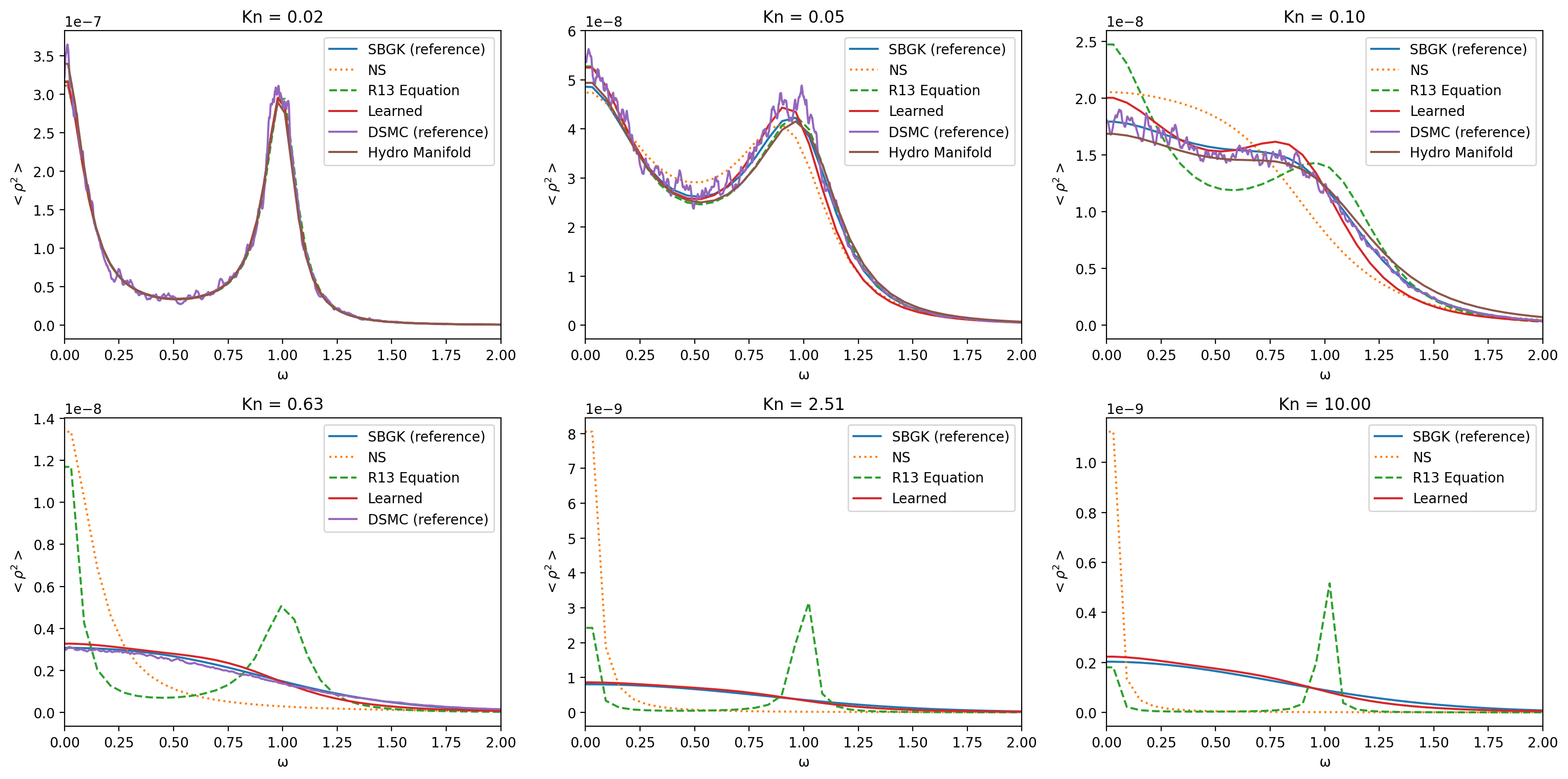}
    \caption{Comparison of density fluctuation curves (horizontal axis temporal frequency $\omega$, vertical axis two-point correlations of density $\langle\rho^2\rangle$): The full Shakhov model (blue), DSMC simulations (purple), the Navier--Stokes equation (orange), the exact spectrally closed hydrodynamics (brown), the R13 model (green) and the learned hydrodynamic manifold (red). In the fully fluidic regime, all models agree well (first row, left and middle). For increasing Knudsen numbers, the Navier--Stokes and the R13 model start to deviate considerably from the kinetic equation and the DSMC simulations, while the learned and the exact spectrally-closed hydrodynamics agree extremely well with the data (first row, right). For larger Knudsen numbers, i.e., wave numbers beyond $k_{c}$, the analytic spectrally-closed hydrodynamics cease to exists and the Navier--Stokes model as well as the R13 model show huge deviations, while the learned optimal hydrodynamics are still in close agreement with the kinetic equation and the DSMC simulation (second row). This demonstrates the optimality of the spectrally closed hydrodynamics and their learned extension.}
    \label{FigComparison}
\end{figure*}

We identify a special lower-dimensional slow manifold as a linear subsystem given by eigenvectors - the \textit{hydrodynamic manifold}  - which attracts all solutions to the overall kinetic equation exponentially fast. The moment closure problem on this special invariant manifold has a unique solution and allows us to define a constitutive law based on this particular closure relation, called the \textit{slow spectral closure} \cite{PhysRevE.110.055105}. By the dynamically optimal attraction properties of the slow manifold, this closure procedure is optimal in the sense that any other closure assumption will be less accurate than the slow spectral closure. In our setting, the optimality of the closure is represented by the proximity of spectral curves as demonstrated strikingly in Figure \ref{FigComparison}.


Salient features of the spectrally closed hydrodynamics are their inherent spatially non-local nature, the existence of a critical wave number and entropy-dissipation balance \cite{kogelbauer2024exact}. As mentioned in the introduction, there are numerous classical closure assumptions on the higher-order fluxes of the kinetic model \eqref{maineq}. These range from Maxwellian closure leading to the Euler equations and the Navier–Stokes–Fourier laws leading to the Navier--Stokes--Fourier equations \cite{chapman1990mathematical}, over Grad's moment closure \cite{grad1949kinetic} and the R13 system \cite{struchtrup2003regularization}, to the maximum entropy principle \cite{levermore1996moment} and quasi-equilibrium projections \cite{gorban2005invariant}. As we demonstrate quantitatively in the subsequent section, all of these existent closure assumptions perform worse when compared to the spectral closure, thus illustrating its optimality.

To construct the slow manifold for the Boltzmann equation \eqref{maineq}, let us first recall its operator spectral properties, see Appendix \ref{AppSpectral} for general spectral properties of kinetic operators. The dynamics on the slow manifold are encoded in the slow eigenvalues of the kinetic operator at each spatial wave number $\bm{k}$,
\begin{equation}\label{eigenvalues}
    -\ri (\bm{k}\cdot v)f + \frac{1}{\rm Kn}Q[f] = \lambda(\bm{k},{\rm Kn}) f,
\end{equation}
There exist five (counted with multiplicity) wave-number dependent eigenvalue branches $\Lambda = \{\lambda_{\rm d}, \lambda_{\rm s},\lambda_{\rm s}, \lambda_{\rm a}, \lambda_{\rm a}^*\}$, comprising the simple real diffusion mode $\lambda_{\rm d}$, the twice-degenerate real shear mode $\lambda_{\rm s}$ and the complex conjugate pair of acoustic modes $(\lambda_{\rm a}, \lambda_{\rm a}^*)$. The five branches of hydrodynamic modes bifurcate from global collision invariants at $\bm{k}=0$ and each one of these modes only exists for a certain range of wave numbers.

The slow manifold for \eqref{maineq} is spanned by the isolated eigenmodes at each wave number and we call the hyperplane spanned by these slow modes the \textit{hydrodynamic manifold},
\begin{equation}\label{defhydromf}
    f_{\rm hydro}(\bm{k},\bm{v}) = {\rm span}\left\{ f_{\lambda}(\bm{k},\bm{v}),\lambda\in\Lambda \right\},
\end{equation}
where $f_\lambda$ is the eigenfunctions corresponding to the eigenvalue $\lambda$, see \cite{PhysRevE.110.055105} as well as Appendices \ref{AppSpectral} and \ref{AppShakhov} for further details. The hyperplane \eqref{defhydromf} acts as a globally exponentially attracting set and allows us to define a dynamically optimal closure. Indeed, the \textit{slow spectral closure} is defined as the unique constitutive law on the hydrodynamic manifold in \eqref{defhydromf} that expresses the higher-order fluxes in terms of the hydrodynamic variables in \eqref{macrovariables}. This, through a linear change of coordinates from spectral to hydrodynamic coordinates, defines a unique linear evolution equation for density, velocity and pressure deviations from equilibrium, which is most easily formulated in frequency space. We emphasize that the dynamic optimality of the spectral closure is a direct consequence of the separation of time scales induced by the ordering of negative real parts in the operator spectrum. We now describe the structure equations of the spectrally closed system on the slow manifold.




In Fourier space, we denote the velocity component along the wave vector $\bm{k}$ as $\hat{u}_{\parallel}$, the longitudinal part, while the two components orthogonal to it are denoted as $(\hat{u}_{\perp,1},\hat{u}_{\perp,2})$, the transversal part, see also Appendix \ref{densityspectra}. We bundle the Fourier transforms of the deviations of the macroscopic variables from the equilibirum state at each wave vector into the vector of hydrodynamic moments,
\begin{equation}\label{defh}
    h = (\hat{\rho}, \hat{u}_{\parallel}, \hat{u}_{\perp,1}, \hat{u}_{\perp,2}, \hat{T} ).
\end{equation}
On the slow manifold, each plane wave component evolves independently according to the spectrally-closed hydrodynamics in frequency space,
\begin{equation}\label{hydro}
    \frac{\partial h}{\partial t} = Th,
\end{equation}
where the entries of the transport matrix $T$,
\begin{equation}\label{defT}
    T = \begin{pmatrix}
0 & -\ri k & 0 & 0 & 0\\
\ri\tau_1(k) & \tau_2(k) & 0 & 0 & \ri\tau_3(k)\\
0 & 0 & \tau_0(k) & 0 & 0\\
0 & 0 & 0 &  \tau_{0}(k) & 0\\
\tau_4(k) & \ri\tau_5(k) & 0 & 0 & \tau_6(k)
    \end{pmatrix},
\end{equation}
are called \textit{generalized transport coefficients}, see \cite{PhysRevE.110.055105} and Appendix  \ref{AppSpectral}.
We stress that, while in classical hydrodynamics, the transport coefficients, such as viscosity or thermal conductivity, are constants, the coefficients in \eqref{defT} are nonlinear scaling laws, i.e., functions of wave number, uniquely defined by the eigenvalues $\Lambda$. In the small wave-number limit, however,
system \eqref{hydro} recovers the linear Euler and the Navier--Stokes equation at zeroth and first order, respectively. Figure \ref{cplot}  in Appendix \ref{AppShakhov} shows the exact $\tau$-curves for the Shakhov model. Note that the dynamics of the transversal components, governed by the shear mode $\tau_0$, decouple from the remaining fields. The matrix \eqref{defT} induces a wave-number dependent modification of entropy which allows to recast the rarefied hydrodynamics in dissipation form, showing that \eqref{hydro} is globally hyperbolic  \cite{PhysRevE.110.055105}. This important feature guarantees that any solution to the rarefied hydrodynamics exists for all times and decays towards the equilibrium.

We emphasize again that the slow spectrally-closed hydrodynamics provide a dynamically optimal solution to the closure problem: any other, generic solution will approach the slow manifold exponentially fast in time and thus minimize the dynamic error for each ensemble of trajectories. Indeed, we show in Section \ref{Results} that the spectrally-closed hydrodynamics outperform any other classical or extended hydrodynamic models, such as Navier--Stokes or regularized thirteen-moments Grad system \cite{karlin1998dynamic,struchtrup2003regularization}, over a large range of Knudsen numbers. In particular, only the slow spectral closure will recover the complete structure of the generalized transport equation \eqref{defT}. The Chapman--Enskog series approximates \eqref{defT} for small wave numbers \cite{kogelbauer2025relation}, while moment projection methods approximate the eigenvalue branches, but might fail to capture essential properties of the eigenvalue branches \cite{karlin2008exact}, such as criticality, as elaborated in the next paragraph. 


However, as mentioned before, each family of frequency-dependent, isolated eigenvectors that spans the hydrodynamic manifold in \eqref{defhydromf} only exists up to a \textit{critical wave number} $k_{c}$, depending on the eigenvalue branch, see Appendices \ref{AppSpectral} and \ref{AppShakhov}. On the one hand, it has been demonstrated that criticality in wave number is an essential feature of the non-local hydrodynamics to capture rarefaction effects \cite{PhysRevE.110.055105}.
On the other hand, criticality implies that the spectrally closed hydrodynamics are only defined for a finite range of wave-numbers. In particular, the numerical evaluation of the analytical spectral closure needs to be adapted for gases at high Knudsen number and we need a practical mechanism to extend the generalized transport equation defined by \eqref{defT} beyond the critical wave number. Furthermore, the exact evaluation of the generalized transport coefficients as predicted by the spectral closure relies on an explicit eigenvalue calculation, which is generally not feasible. To overcome the limitation of criticality and the restrictions of the spectral quantities involved, we will use the structural properties of the transport matrix as predicted from the slow closure to learn the transport coefficients from density fluctuations as detailed in the next section.\\



\section{Learning Optimal Hydrodynamics from Density Fluctuations}
\label{LearningHydroSec}

The linear dynamics in \eqref{hydro} with frequency-dependent transport matrix as defined in \eqref{defT} govern the time-evolution of optimal hydrodynamics. We will now use the structure of the transport matrix to learn these optimal constitutive laws from light-scattering data. As discussed before, this allows us to bypass the explicit calculation of eigenvalues and extend the optimal slow hydrodynamic beyond the critical wave number. As a practical case to illustrate the optimality of the spectral closure, we compare results of the light-scattering experiment - a well established experimental test case for the rarefied gas flows \cite{miles2001laser,wang2021rayleigh}, for which the linear kinetic description \eqref{maineq} is sufficient.



To demonstrate that the generalized transport matrix can indeed predict the dynamics of gases over the full range of Knudsen numbers, we generate a reference DSMC and Shakhov kinetic data set. Throughout, we define the Knudsen number for wave number $k$ as
\begin{equation} \label{Kn_k}
   \text{Kn} = \frac{k\mu }{ 2 \pi\rho} \sqrt{\frac{m}{k_B T}}, 
\end{equation}
where $\mu$ is the dynamic viscosity, $\rho$ is the reference density, $m$ is the molecular mass, $T$ is the reference temperature and $k_B$ is Boltzmann's constant. For details of the DSMC computations, we refer to Appendix \ref{AppDSMC}. 

As an underlying benchmark kinetic model, we use the linear Shakhov collision operator \cite{shakhov1968generalization} in the following. On the one hand, the Shakhov model is complex enough to allow for Prandtl numbers $0\leq \rm{Pr} \leq 1$, including the physically relevant $\rm{Pr}=2/3$ and shows excellent agreement with full DSMC simulations, see, e.g., \cite{ambrucs2020comparison} for a comparative study, as well as the direct comparison in Figure \ref{FigComparison}. On the other hand, the Shakhov collision kernel is still simple enough to carry out an explicit operator spectral analysis and give closed-form expressions for the spectral closure \cite{KOGELBAUER2024134014}. In particular, the Shakhov model comprises the widely-used BGK model for $\rm Pr = 1$ \cite{bhatnagar1954model} as a special case. 
Analogously as for the three-dimensional BGK operator \cite{PhysRevE.110.055105},
we calculate the spectral closure for the Shakhov operator explicitly for the first time in this work, see Appendix \ref{AppShakhov}. The critical Knudsen number 
\begin{equation}\label{Kncrit}
\rm Kn_{\rm crit} = 0.1517,
\end{equation}
is derived from the critical wave number, see Appendix \ref{AppShakhov} for the physical parameter values of Table \ref{AP_tab:table2}. This will serve as a first test for the accuracy of our learning algorithm below the critical wave number as discussed in the next section in detail. Furthermore, we carry out a direct numerical simulation of the Shakhov model based on a Monte Carlo scheme, as detailed in Appendix \ref{AppShakhov} as well. To avoid confusion in the terminology, we emphasize the difference between the operator spectra, which are the basis for the exact spectral closure, and the light scattering spectra, which are used to train a neural network.



To relate the generalized transport coefficients to macroscopic measurements, especially stochastic fluctuations around equilibrium, we recall the light-scattering experiment. Rayleigh--Brillouin scattering is a classical and well-established technique to determine fluctuations in macroscopic observables from changes in the fluid's dielectric field \cite{hansen2013theory}. To this end, a fluid is probed by an incident electromagnetic wave with frequency $\omega_{\rm inc}$ at a spatial wave vector $\bm{k}_{\rm inc}$. The intensity of the scattered wave $I(\omega, \bm{k})$ is then measured against the frequency shift $\omega$ and the wave vector shift $\bm{k}$. The two-point correlation function of deviations in density fluctuations can then be recovered according to $   I(\omega, \bm{k}) \propto \langle \rho^2 \rangle ( \omega, \bm{k})$. For further physical details, we refer to Appendix \ref{AppLightScattering}.

The structure of the hydrodynamic equations \eqref{defT} in combination with density-fluctuation data allows us to introduce a parametric, low-dimensional learning algorithm for the generalized transport coefficients. Learning scaling laws from data defines a non-convex optimization problem, which strongly suggests the use of neural networks and stochastic optimization techniques \cite{goodfellow2016deep}. The learning data is given by the density, velocity and temperature fluctuations, derived either from DSMC data or from the Shakhov model directly. The learned density fluctuation curve is an analytical function of the generalized transport coefficients, which are modeled as outputs of a neural network, see Figure \ref{neuralnet}. We constrain their values and derivatives at $k=0$ to guarantee consistency with the Navier-Stokes equation for small wave numbers.
We use the ADAM \cite{kingma2014adam}
optimizer to minimize the loss function and determine the weight vectors. We refer to Appendix \ref{AppNeuralNet} and \ref{AppLearning} for details on the implementation of the learning algorithm. 

Our approach fundamentally differs from methods that learn the time evolution of moments, such as e.\ g. \cite{Hana2019}, since we do not use trajectory data as an input, but much rather learn the generalized transport coefficients in Fourier space directly. The theoretical insight of the spectral closure and the resulting structural properties of the generalized transport matrix (\ref{defT}) thus offer a clear advantage over PINN-related approaches. Indeed, methods focusing on trajectories inherently face an exponential growth of errors over time, thus restricting their applicability to short time intervals. Our approach, however, relying on the parameterization of transport coefficients, captures the full temporal spectrum and yields bounded errors that remain stable over arbitrarily long timescales thanks to the hyperbolicity guaranteed by entropy-dissipation balance of the optimal hydrodynamics. This feature is essential for accurate and efficient long-time predictions in rarefied gas dynamics.

To ensure numerical stability and physical fidelity of the learned constitutive relations, we employ a subset of the generalized transport coefficients rather than the complete set. This constraint mitigates instability while preserving the model's ability to generalize across the Knudsen number range under consideration. Our training dataset comprises 2048 carefully sampled spectra within the Knudsen number range of 0 to 10, with inference performed within the same domain, thereby avoiding extrapolation and its associated risks, see Appendix \ref{AppNeuralNet}. 
We emphasize that extrapolation is never reliable, independent of whether neural networks are used as a predictive routine. Much rather, the spectral closure allows us  to define a low-parametric, functional learning problem for the generalized transport coefficients. The Shakhov model and the DSMC data are used to learn these functions over a wide range of Knudsen numbers (Kn 0 to 10), which covers most of the practically relevant rarefaction regimes. Further extensions are, of course, easiliy possible. 



\begin{figure*}
    \centering
\includegraphics[width= 17.8cm]{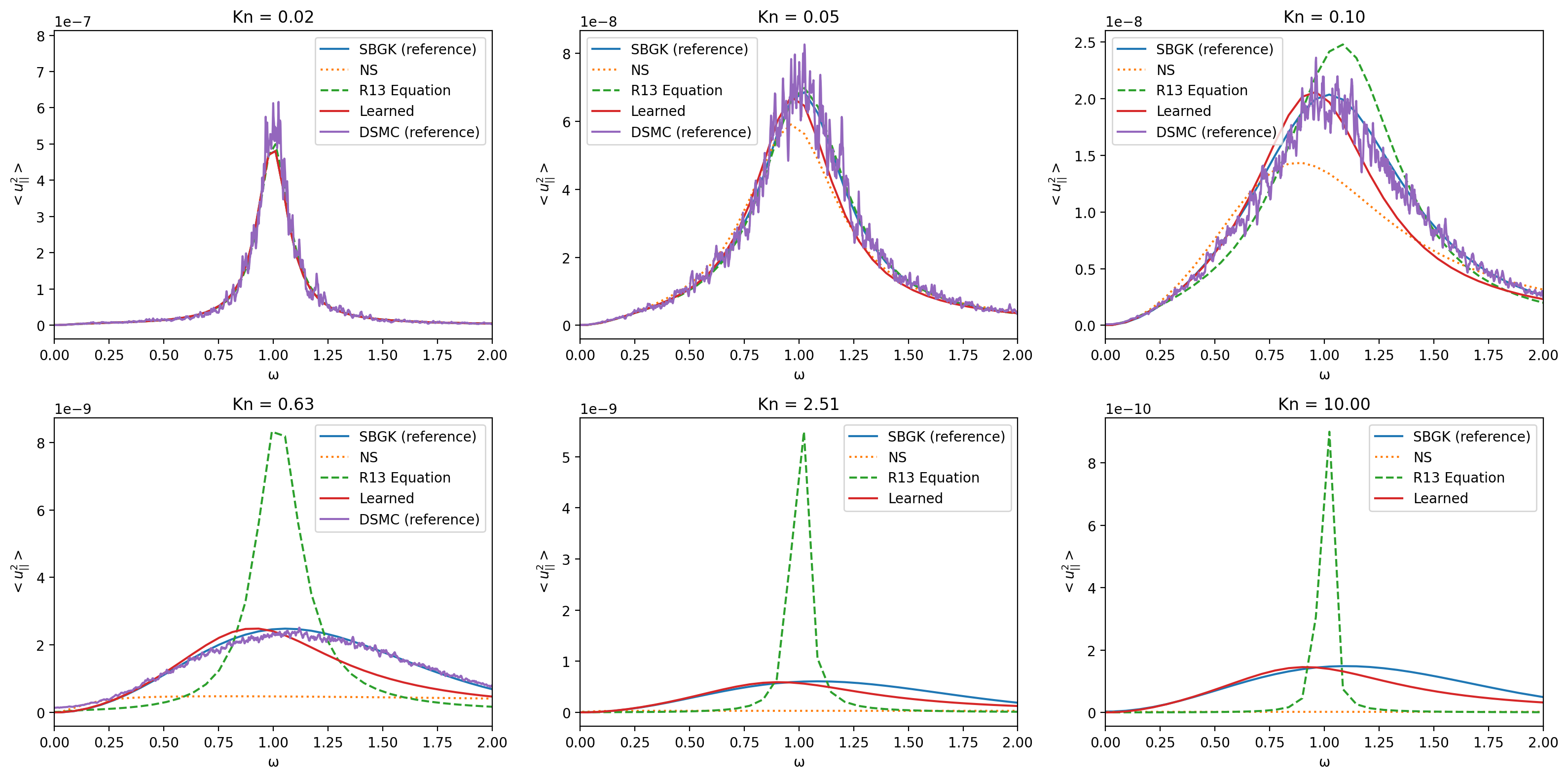}
    \caption{Comparison of velocity fluctuation curves (horizontal axis temporal frequency $\omega$, vertical axis intensity $I$): The full Shakhov model (blue), DSMC simulations (purple), the Navier--Stokes equation (orange), the exact spectrally closed hydrodynamics (brown), the R13 model (green) and the learned hydrodynamic manifold (red). In the fully fluidic regime, all models agree well (first row, left and middle). For increasing Knudsen numbers, the Navier--Stokes and the R13 model start to deviate considerably from the kinetic equation and the DSMC simulations, while the learned and the exact spectrally-closed hydrodynamics agree extremely well with the data (first row, right). For larger Knudsen numbers, i.e., wave numbers beyond $k_{c}$, the analytic spectrally-closed hydrodynamics cease to exists and the Navier--Stokes model as well as the R13 model show huge deviations, while the learned optimal hydrodynamics are still in close agreement with the kinetic equation and the DSMC simulation (second row).}
    \label{FigComparisonV}
\end{figure*}

Furthermore, the ability of the model to accurately predict velocity fluctuation spectra — a task different from the training objective — provides additional evidence against overfitting and supports the robustness of the learned transport coefficients, see Figure \ref{FigComparisonV}. Together, these results highlight the capability of our approach to achieve stable, accurate, and interpretable predictions over long time horizons, distinguishing it from existing methods.




\section{Main Results: Comparison to DSMC data, the Shakhov model and Extended Hydrodynamics}
\label{Results}

\begin{figure}
    \centering
\includegraphics[width= 9cm]{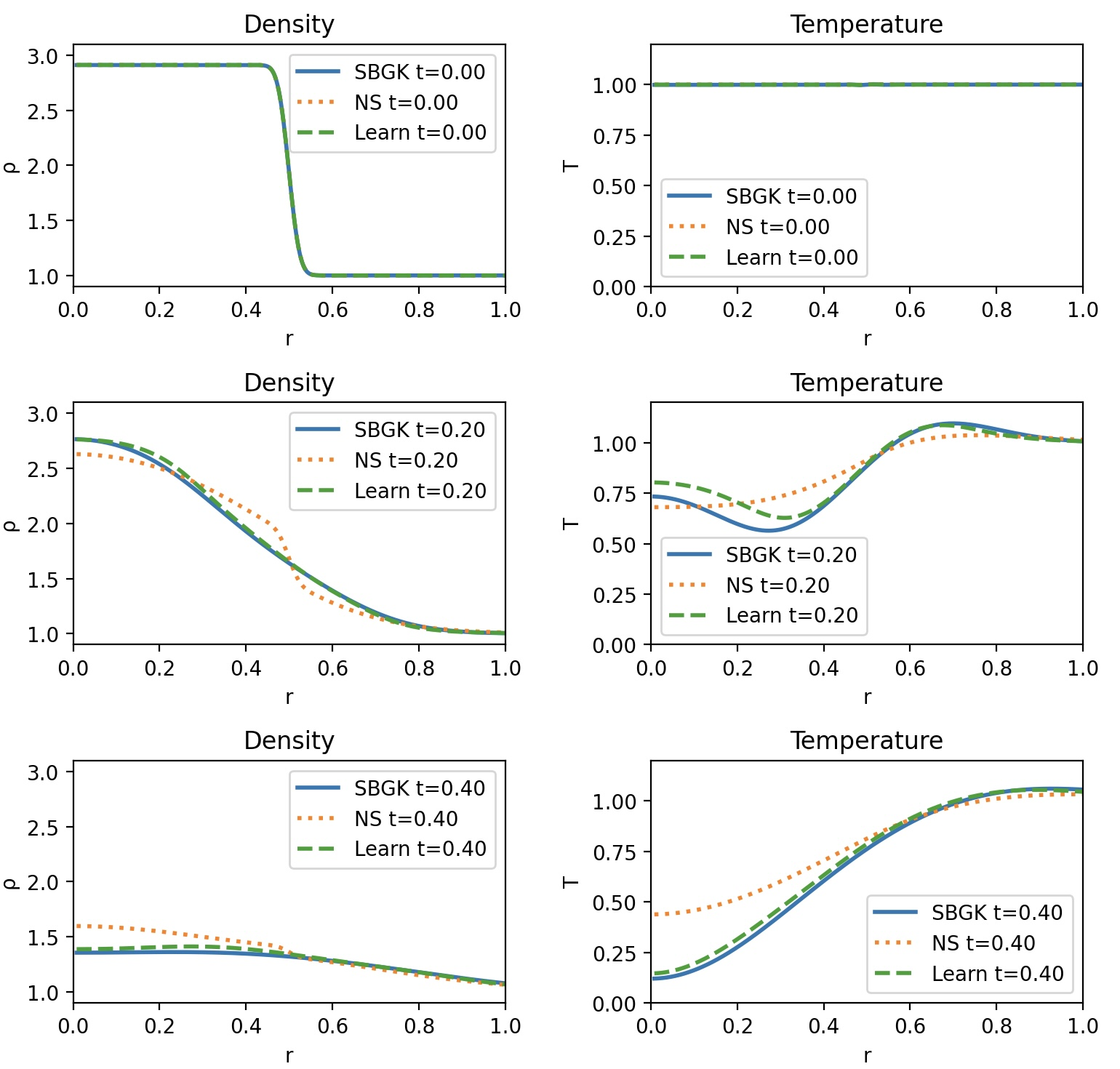}
    \caption{Comparison of moment advection for density and temperature according to the generalized transport matrix in  \eqref{defT} for the parameter values in Table \ref{AP_tab:table2}.
    An initial, radially-symmetric density jump at constant temperature evolves over time according to the full Shakhov model (blue), the learned hydrodynamics (green) and the Navier--Stokes equation (orange).
    }
    \label{Figadvection}
\end{figure}

With all the ingredients at hand, we are now ready to present and discuss our main results shown in Figure \ref{FigComparison}, Figure \ref{FigComparisonV} and Figure \ref{Figadvection}. We compare the density fluctuation spectra of the exact and learned non-local hydrodynamics to the full Shakhov model and to three-dimensional DSMC data. Both for the comparison to the training data in \ref{FigComparison} as well as for the comparison to the out-of-sample test, the velocity fluctuations in \ref{FigComparisonV}, we only show the positive semi-axis since the light-scattering spectra are symmetric with respect to $\omega$.  To this end, we simulate the full Shakhov equation via a Monte Carlo method, see Appendix \ref{AppShakhov}, for different values of Knudsen number (purple curves in Figure \ref{FigComparison}). As mentioned before, we perform three-dimensional DSMC simulations
as input data for our learning scheme as well as for comparison to the analytically obtained spectrally closed hydrodynamics. As a first consistency check, we note that the Shakhov model agrees extremely well with the DSMC result for all Knudsen numbers, consistent with previous observations \cite{ambrucs2020comparison}. We emphasize that the Shakhov model and the DSMC data are the underlying kinetic model from which the spectral closure is obstinate, i.e., the grand truth, to which we compare the learned constitutive laws. A closure relation is thus accurate, if it agrees well with both the DSMC and the Shakhov data.


In addition to a direct comparison to the Shakhov kinetic model and DSMC data, we contrast the present non-local hydrodynamics to other higher-order hydrodynamic models. Firstly, we show the density-fluctuation curves for the Navier--Stokes equation (orange curve in Figure \ref{FigComparison}) as the classical fluid dynamics model. Secondly, we simulate the R13 extended hydrodynamics \cite{karlin1998dynamic,struchtrup2003regularization}, a widely used model for moderately rarefied gas flows (green curve in  Figure \ref{FigComparison}). We emphasize that the density fluctuation curves in Figure \ref{FigComparison}
are shown in physical units for a certain set of parameter values. The analytical spectral closure derived from the Shakhov model (brown curve in  Figure \ref{FigComparison}) shows excellent agreement with the full kinetic model and the DSMC data up to 
the critical Knudsen number \eqref{Kncrit}.



While all extended hydrodynamic models agree well with the kinetic and the DSMC data for small Knudsen numbers, the analytic spectrally closed hydrodynamics outperforms the Navier--Stokes equation and the R13 model decisively for higher Knudsen numbers (pronounced deviations at around Kn$=0.1$), all the way up to the critical wave number. 
While we report here only the R13 as a representative example of higher-order hydrodynamics, many more models were compared recently by Wu and Gu \cite{wu2020accuracy}, with the conclusion that none of them are able to predict the spectra accurately for ${\rm Kn}\geq 0.05$. 
Our result proves the dynamical optimality of the spectral closure: any other closure procedure will necessarily have a larger deviation from the underlying kinetic dynamics within its domain of existence.


The learned hydrodynamic closure (red curve in Figure \ref{FigComparison}) is also in close agreement with the both the Shakhov as well as the DSMC data. The analytic closure serves as a further benchmark to ensure the accuracy of the learned closure up to the cirtical wave number. While the analytical closure is limited to frequencies below the critical wave number, the learned hydrodynamics extends to arbitrarily large wave numbers
in good agreement with the data (second row in Figure \ref{FigComparison}). At ${\rm Kn}=10$, we observe slight deviations of the learned curves due to boundary effects of the learning algorithm. As an out-of-sample test, we apply the transport curves learned from density fluctuations to velocity fluctuations, as shown in Figure \ref{FigComparisonV}. Again, the the learned curves show excellent agreement with the Shakhov data, while the R13 and Navier--Stokes spectra show huge deviations for larger Knudsen numbers.

As a further illustration of the optimality of the spectrally closed hydrodynamics, we compute the time evolution of density and temperature directly.
Figure \ref{Figadvection} shows the advection of a sharp, radially symmetric density drop at constant initial temperature on the whole space, see Appendix \ref{AppTimeEvolution} for details on the numerical implementation and the spatial Fourier transform of the initial condition, indicating that the density drop contains frequencies beyond the critical wave number. 

The learned non-local hydrodynamics (green) show almost perfect agreement with the Shakhov model (blue) over the whole transient regime, while the Navier--Stokes solution (orange), deviates considerably from the true solution, especially towards the center. Rarefaction effects become more pronounced for larger spatial wave numbers in the Fourier transform of density and temperature. Since the initial density profile has a rather sharp interface, the higher frequencies can be resolved with the non-local hydrodynamics, while the Navier--Stokes equation deviates more from the full Shakhov model in this transient time interval. For larger times, higher frequencies dissipate faster and the Navier--Stokes equation provides and equally good fit as compared to the non-local hydrodynamics. We stress again at this point that the learned rarefied hydrodynamics can be extended for arbitrarily long times thanks to the entropy-dissipation principle \cite{PhysRevE.110.055105}, while agreeing extremely well with the underlying kinetic model in the transition regime.

\section{Discussion}
\label{DiscussionSec}

We presented the dynamically optimal solution to the moment closure problem for linear kinetic equations. Based on the theory of slow manifolds and spectral insights, we derived a general closure framework leading to novel, non-local hydrodynamics for a large range of Knudsen numbers. While the analytical optimal hydrodynamics are limited by criticality in wave number, the generalized transport matrix based on spectral quantities provides a framework for extended hydrodynamics for arbitrarily large wave numbers. Within this framework, we learned the generalized transport coefficients from density-fluctuation data. Our results show excellent agreement with DSMC simulations and the Shakhov kinetic equation up to $\rm{Kn}=\mathcal{O}(10)$.

We therefore addressed two issues of rigorous hydrodynamics: the dependence on spectral information and the limitations of criticality. While the explicit evaluation of generalized transport coefficients is feasible for certain kinetic models such as the BGK or the  Shakhov equation, quantitative spectral data is scarce for Boltzmann's hard-sphere collision integrals. Furthermore, 
rigorous hydrodynamics are non-local in space and the question arises how to uniquely extend it beyond the critical wave number. Both questions were addressed in this work using the learning of rigorous constitutive relation. 

We showed hat the analytically computed hydrodynamics match the the full Shakhov equation, thus demonstrating that rigorous hydrodynamics are indeed optimal (in the sense that they outperform any other hydrodynamic closure) up to the critical wave number. Furthermore, the learned constitutive equations, while as accurate as the analytical spectral closure up to the critical wave number, extend way beyond the critical wave number while maintaining the same excellent accuracy.
The learned transport coefficients perform accurately in other multi-scale settings as well, as demonstrated by a sharp density propagation setup. Thus, the application of machine learning to circumvent a cumbersome spectral problem and to extend constitutive relations beyond criticality solves the problem of optimal hydrodynamics near equilibrium independent of the degree of rarefaction. As a consequence, the dynamical optimality of the spectral closure as a solution to Hilbert's sixth problem on the passage from kinetic theory to hydrodynamics has been validated on the linear level empirically as well. 






\section{Acknowledgement}
F.K and I.K. were supported by the European Research Council (ERC) Advanced Grant 834763-PonD. Computational resources at the Swiss National Super Computing  Center  CSCS  were  provided  under the grant s1286.

\bibliography{Learning_Transport_Coef_Bib}

\clearpage

\appendix

\section{General Spectral Properties of Kinetic Operators and Spectral Closure}
\label{AppSpectral}

In this appendix, we recall the basic spectral properties of kinetic operators in frequency space, 
\begin{equation}\label{kinoperator}
\mathcal{L}_{\bm{k}} = -\ri\bm{v}\cdot\bm{k} + Q,
\end{equation}
which are obtained from \eqref{maineq} by Fourier transform and allowing us to treat each plane wave contribution separately. Furthermore, we recall the general theory of spectral closure as derived in \cite{PhysRevE.110.055105}. 

\subsection*{Spectral Theory}
The spectrum of a Boltzmann-type linear operator, denoted as $\sigma(\mathcal{L}_{\bm{k}})$ henceforth, is stable thanks to entropy increase, i.e., 
\begin{equation}
  \Re\sigma(\mathcal{L}_{\bm{k}})\leq 0, 
\end{equation}
and eigenvectors with zero real part correspond to global collision invariants.
The operator spectrum $\sigma(\mathcal{L}_{\bm{k}})$ splits into an essential part $\sigma_{\rm ess}(\mathcal{L}_{\bm{k}})$, corresponding to fluctuations, and discrete, isolated eigenvalue branches, parameterized by spatial wave number, corresponding to hydrodynamics \cite{ellis1975first,palczewski1984}:
\begin{equation}
\sigma(\mathcal{L}_{\bm{k}}) = \sigma_{\rm ess}(\mathcal{L}_{\bm{k}}) \cup \sigma_{\rm disc}(\mathcal{L}_{\bm{k}}).
\end{equation}The eigenvalue branches bifurcate from collision invariants and only exists up to a critical wave number, thus limiting the range of existence of hydrodynamics. This phenomenon of criticality is vital for the representation of rarefaction effects on the macroscopic level \cite{PhysRevE.110.055105}.\\
Since the hydrodynamic branches lie above the essential spectrum, their corresponding eigenvectors span a slow manifold (hyperplane) and a general trajectory will approach this particular manifold - and only this manifold - exponentially fast. The existence of the slow manifold through spectral properties thus induces a hierarchy of time scales, first by the splitting between discrete and essential spectrum, and then among eigenvalues ordered by the magnitude of their negative real parts. The separation strength in turn scales with Knudsen number, proving that in the classical fluid limit $\rm{Kn}\to 0$, \textit{all} dynamics are governed by hydrodynamic modes, while in the ballistic limit $\rm{Kn}\to\infty$, \textit{all} dynamics are governing by fluctuations corresponding to the essential spectrum.

\subsection*{Spectral Closure}
In this subsection, we recall general properties of the spectral closure technique as detailed in \cite{PhysRevE.110.055105}. As discussed before, for each wave number $k$, the primary hydrodynamic spectrum consists of four branches of eigenvalues,
\begin{equation}\label{LambdaHydro}
    \Lambda_{\rm hydro}(k) = \{\lambda_{\rm s}(k), \lambda_{\rm d}(k),\lambda_{\rm a}(k),\lambda_{\rm a}^{*}(k)\},
\end{equation}
where in the limit $k\to 0$, the four branches in \eqref{LambdaHydro} collapse to zero, which corresponds to the collision invariants of the linear part (center modes). Here, $\lambda_{\rm s}$ denotes the double degenerate, semi-simple, real shear mode, $\lambda_{\rm d}$ is the real diffusion mode and $(\lambda_{\rm a},\lambda_{\rm a}^*)$ is the pair of complex conjugated acoustic modes. 

For any $\bm{k}$, let $\bm{Q}_{\bm{k}}$ denote the unique rotation such that $\bm{k}=\bm{Q}_{\bm{k}}(k,0,0)$ and define the block-diagonal $5\times 5$ matrix,
\begin{equation}\label{eq:Qtilde}
\tilde{{Q}}_{\bm{k}} = \diag(1,\bm{Q}_{\bm{k}},1).
\end{equation}

Denoting the normalized hydrodynamic moments as
\begin{equation}
  e  = \left(1,\bm{v},\frac{{v}^2-3}{\sqrt{6}}\right),
\end{equation}
the spectral closure can be derived from the the $5\times 5$ \emph{spectral matrix},
\begin{equation}\label{GLP}
{{G}(\lambda,\bm{k})} = (2\pi)^{-\frac{3}{2}}\int_{\mathbb{R}^3}{e}\otimes\Big((\mathcal{L}_{\bm{k}}+\mathbb{P}_5-\lambda)^{-1}{e}\Big)\exp\left({-\frac{{v}^2}{2}}\right)d\bm{v},
\end{equation}
where $\mathbb{P}_5$ denotes the projection onto the hydrodynamic moments.

We define the spectral closure \cite{PhysRevE.110.055105} for to the five-dimensional slow eigenspace,
\begin{equation}
    F_{\bm{\lambda}} = [\hat{f}_{\lambda_{\rm d}},\hat{f}_{\lambda_{\rm a}},\hat{f}_{\lambda_{\rm a}^*},\hat{f}_{\lambda_{\rm s},1},\hat{f}_{\lambda_{\rm s},2}]
\end{equation} 
associated to \eqref{LambdaHydro}, as the unique linear operator $\mathcal{C}_{\rm spectral}:\text{range }\mathbb{P}_5\to \text{range }\mathbb{P}_5^{\perp}$, through the relation,
\begin{equation}\label{specclosure}
    \mathcal{C}_{\rm spectral} h  = \mathbb{P}_5^{\perp}F_{\lambda}H_{\bm{k}}^{-1} h,
\end{equation}
i.e., the macroscopic variables are expressed in spectral coordinates, projected onto the slow manifold associated to the hydrodynamic modes and then projected onto the orthogonal complement of the collision invariants. The coordinate change from spectral to physical coordinates is given by \begin{equation}\label{defHfinal}
     H_{\bm{k}} = \tilde{Q}_{\bm{k}} \begin{pmatrix}
         1 & 1 & 1 & 0 & 0\\
         \frac{\ri }{k}\lambda_{\rm d} &  \frac{\ri}{k}\lambda_{\rm a} &  \frac{\ri}{k}\lambda_{\rm a}^* & 0 & 0\\
         0 & 0 & 0 & 1 & 0\\
         0 & 0 & 0 & 0 & 1\\
         \theta(\lambda_{\rm d}) &  \theta(\lambda_{\rm a}) &  \theta(\lambda_{\rm a}^*) & 0 & 0
     \end{pmatrix}. 
 \end{equation}

Here, $\theta$ denotes the \textit{spectral temperature}, given by a quotient of Riesz projections,
\begin{equation}\label{eq:spectemp}
   {\theta(\lambda,\bm{k})} = \frac{\text{adj}[{G(\lambda,\bm{k})}-I]_{1,5}}{\text{adj}[{G(\lambda,\bm{k})}-I]_{1,1}}.
\end{equation}

The spectrally-closed hydrodynamic system is given by 
\begin{equation}\label{dynh}
    \frac{\partial h}{\partial t } = Th,\quad  T = H_{\bm{k}}\Lambda H_{\bm{k}}^{-1},
\end{equation}
where
\begin{equation}
    h = (\hat{\rho},\hat{u}_{\parallel},\hat{u}_{\perp 1},\hat{u}_{\perp 2},\hat{T})
\end{equation}
denotes the hydrodynamic variables in $\bm{Q}_{\bm{k}}$-coordinates, the diagonal matrix $\Lambda$ contains the primary hydrodynamic eigenvalues in \eqref{LambdaHydro} and the matrix $H_{\bm{k}}$ realizes the coordinate change from spectral to physical coordinates at each wave number.\\
As shown in \cite{PhysRevE.110.055105} the general form of the transport matrix is given by
\begin{equation}\label{defTShakhov}
    T = \begin{pmatrix}
0 & -\ri k & 0 & 0 & 0\\
\ri \tau_1 &  \tau_2 & 0 & 0 & \ri \tau_3\\
0 & 0 & \tau_0 & 0 & 0\\
0 & 0 & 0 & \tau_0 & 0\\
\tau_4 & \ri \tau_5 & 0 & 0 & \tau_6
    \end{pmatrix},
\end{equation}
where the detailed form of the (non-local) transport coefficients is given by the formula
\begin{equation}\label{transportcoef}
    \begin{split}
&\tau_0=\lambda_s,\\
     %
&     \tau_1  =\frac{2}{k^2\det{H}}\left[ \lambda_{\rm d}\Im[\lambda_{\rm a}^*(\lambda_{\rm d}-\lambda_{\rm a})\theta(\lambda_{\rm a})]-|\lambda_{\rm a}|^2(\Im\lambda_{\rm a})\theta(\lambda_{\rm d})  \right],\\
     %
      %
&      \tau_2  =  \frac{2}{k\det{H}}\left[2(\Re\lambda_{\rm a})(\Im\lambda_{\rm a})\theta(\lambda_{\rm d})-\Im[(\lambda_{\rm d}^2-(\lambda_{\rm a}^*)^2)\theta(\lambda_{\rm a})]\right],\\
      %
      %
&      \tau_3  = \frac{\sqrt{6}}{k^2\det{H}}|\lambda_{\rm d}-\lambda_{\rm a}|^2(\Im\lambda_{\rm a}),\\
  %
%
& \tau_4  = \frac{\sqrt{6}}{k\det{H}} \left[\Im[\lambda_{\rm a}(\lambda_{\rm d}-\lambda_{\rm a})\theta(\lambda_{\rm a})\theta(\lambda_{\rm d})]+\lambda_{\rm d}(\Im\lambda_{\rm a})|\theta(\lambda_{\rm a})|^2  \right],\\
%
%
& \tau_5  = \frac{\sqrt{6}}{\det{H}} \left[\theta(\lambda_{\rm d})\Im[\theta(\lambda_{a})(\lambda_{\rm d}-\lambda_{\rm a})]+(\Im\lambda_{\rm a})|\theta(\lambda_{a})|^2 \right],\\
%
%
& \tau_6   = \frac{2}{k\det{H}} \left[\lambda_{\rm d}\theta(\lambda_{\rm d})(\Im\lambda_{\rm a})+\Im[\lambda_{\rm a}\theta(\lambda_{\rm a})(\lambda_{\rm a}^*-\lambda_{\rm d})] \right],  \\
&\det{H} = \frac{2}{k}\left((\Im\lambda_{\rm a})\theta(\lambda_{\rm d})-\Im[(\lambda_{\rm a} - \lambda_{\rm d})\theta(\lambda_{\rm a}^*)]\right).    
    \end{split}
    \end{equation}

\section{Rayleigh-Brillouin Scattering and Calculation of Fluctuation Spectra}
\label{AppLightScattering}

In this appendix, we recall the connection between light-scattering and density fluctuations in a fluid. First, we recall some basic definitions and results from harmonic analysis. We sketch the basic physical mechanism of the light-scattering experiment and derive the relation between variations in the dielectric field and the two-point correlation of density fluctuations.  Then, we describe how the two-point density fluctuations can be recovered from light-scattering data.\\

\subsection*{Definition of Averages and Fourier Transforms}\label{AppFourier}

The spatial Fourier transform of an integrable function $f:\mathbb{R}^3\to\mathbb{C}$ is denoted as
\begin{equation}\label{defFouriertransform}
    \hat{f}(\bm{k}) = \frac{1}{(2\pi)^{3/2}} \int_{\mathbb{R}^3} f(\bm{x}) e ^{-\ri \bm{k}\cdot\bm{x}}\,  d\bm{x},
\end{equation}
while the one-sided (temporal) Fourier transform of an integrable function $f:[0,\infty)\to\mathbb{C}$ is denoted as 
\begin{equation}\label{defonesidedFourier}
   f^{+}(\omega) = \frac{1}{\sqrt{2\pi}}\int_0^{\infty}  f(t) e^{-\ri \omega t}\, dt, 
\end{equation}
For a real-valued function $f$ with period $T$, we denote its average as
\begin{equation}
    \langle f \rangle_T = \frac{1}{T} \int_{-\frac{T}{2}}^{\frac{T}{2}}   f(s)\, ds
\end{equation}
For any two $T$-periodic functions $f,g$, their temporal correlation is given as
\begin{equation}\label{cor}
     \langle f g\rangle (t) = \langle f(s) g(s+t) \rangle_T.
\end{equation}
In particular, the two-point correlation function of a $T$-periodic function is given by 
\begin{equation}\label{twopoint}
    \langle f^2\rangle (t) = \langle f(s) f(s+t) \rangle_T.
\end{equation}
Clearly, the correlation between periodic functions commutes with taking any (spatial/temporal) derivative,
\begin{equation}\label{commder}
    \langle f \partial g\rangle = \partial \langle f g\rangle.
\end{equation}
From the reflectional symmetry of the integrals in \eqref{twopoint}, we deduce that $  \langle f^2\rangle (-t) =   \langle f^2\rangle (t)$.\\
Let es denote the Fourier transform of a $T$-periodic function as 
\begin{equation}
    \mathcal{F}_t[f](\omega) = \frac{1}{\sqrt{2\pi}}\int_{-\frac{T}{2}}^{\frac{T}{2}}  f(t) e^{-\ri \omega t}\,  dt. 
\end{equation}
The two-point correlation in \eqref{twopoint} in frequency space reads
\begin{equation}
\begin{split}
        \langle f^2 \rangle (\omega) & = \langle \mathcal{F}_t[f(s)f(s+t)] \rangle_T\\
        & = \frac{1}{\sqrt{2\pi}T} \int_{-\frac{T}{2}}^{\frac{T}{2}} \int_{-\frac{T}{2}}^{\frac{T}{2}}f(s) f (s+t) e^{-\ri \omega t} dtds, 
\end{split}
\end{equation}
and relates to the intensity of the Fourier transform through the Wiener--Khinchin theorem \cite{wiener1930generalized},
\begin{equation}\label{WienerKinchin}
        \langle f^2\rangle(\omega) = \frac{\sqrt{2\pi}}{T} |\mathcal{F}_t(f)(\omega)|^2. 
\end{equation}
\eqref{WienerKinchin} naturally extends to vector-valued functions.
According to \eqref{defonesidedFourier}, the one-sided Fourier transform of \eqref{twopoint} is given by
\begin{equation}
\label{onesided}	
   \langle f^2\rangle^{+}(\omega) = \frac{1}{\sqrt{2\pi}}\int_0^{\infty}   \langle f^2\rangle_T (t) e^{-i \omega t}\, dt. 
\end{equation}
The full temporal Fourier transform
\begin{equation}
    \widetilde{\langle f^2 \rangle} (\omega) = \frac{1}{\sqrt{2\pi}} \int_{-\infty}^\infty   \langle f^2 \rangle_T(t)  e^{-\ri \omega t }\, dt,
\end{equation}
can be recovered from \eqref{onesided} via the formula
\begin{equation}
     \widetilde{\langle f^2 \rangle} = 2 \Re[\langle f^2\rangle^{+}]. 
\end{equation}
The one-sided Fourier transform satisfies
\begin{equation}
    (\partial_t f)^{+}(\omega) = \ri \omega f^{+}(\omega) -\frac{1}{\sqrt{2\pi}} f(0) ,
\end{equation}
for any function $f$ with suitable decay at infinity.

\subsection*{Rayleigh--Brillouin Scattering}
\label{RBScattering}

\begin{figure}[h!]
    \centering
    \includegraphics[width=1\linewidth]{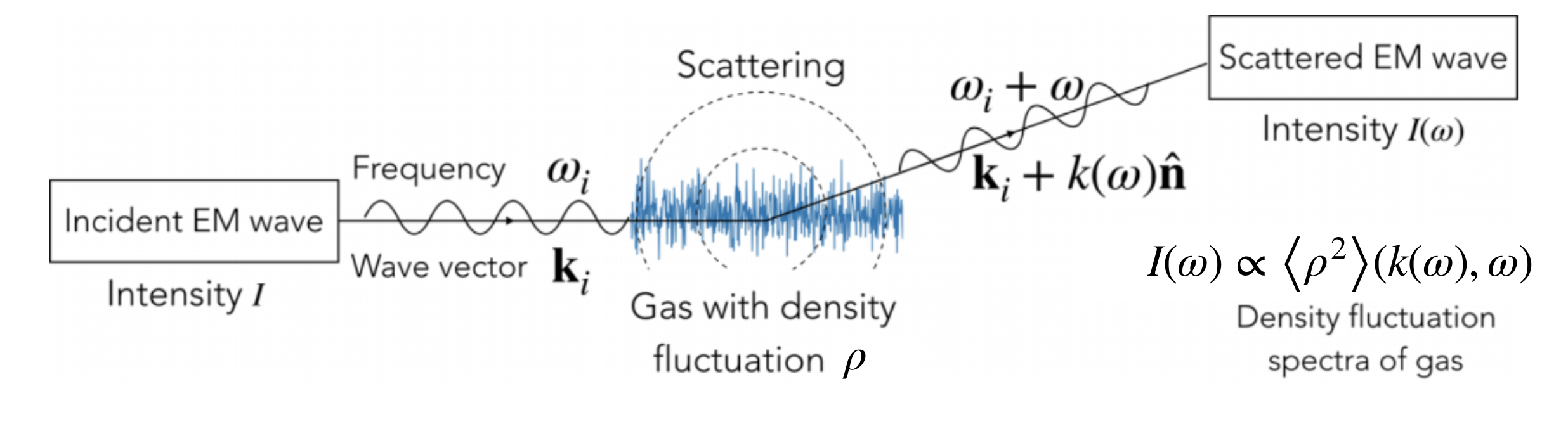}
    \caption{Schematics of the light-scattering experiment: An incident electromagnetic wave with temporal frequency $\omega_{\rm inc}$ and spatial wave vector $\bm{k}_{\rm inc}$ is scattered by a fluid with density fluctuations $\rho$. The intensity $I(\omega)$, depending on the frequency shift $\omega$, defines the Rayleigh--Brillouin spectra, which are proportional to the two-point correlation function of the density fluctuations. 
}
    \label{Lightscatterinsetup}
\end{figure}

In this section, we recall the basic physical mechanisms of the light-scattering experiment. For an in-depth discussion of Rayleigh--Brillouin scattering, we refer to \cite{landau2013electrodynamics}. Rayleigh--Brillouin scattering relates the refraction of an incident electromagnetic wave scattered by a fluid to its stochastic density fluctuations $\rho$, see Figure \ref{Lightscatterinsetup}. The incident signal is a plane wave,
\begin{equation}\label{emwave}
 \bm{E}_{\rm inc}(\bm{x},t) = \bm{\xi}_0 \exp(\ri \bm{k}_{\rm inc}\cdot \bm{x} + \ri \omega_{\rm inc} t),
\end{equation}
where $\bm{\xi}_0$ is the constant polarization vector, $\bm{k}_{\rm inc}$ is the spatial wave vector of the incidence wave and $\omega_{\rm inc}$ is the corresponding temporal frequency. The spatial and temporal frequencies are related through
\begin{equation}
    |\bm{k}_{\rm inc}| = \frac{\sqrt{\varepsilon_0} \omega_{\rm inc}}{c},
\end{equation}
where $c$ is the speed of light and $\varepsilon_0$ is the unperturbed dielectric constant of the gas. The electromagnetic wave in \eqref{emwave} is a solution to Maxwell's equation in matter (assuming that the permeability of the fluid is the same as for the vacuum), 
\begin{equation}\label{Maxwell}
    \begin{split}
        \nabla\cdot\bm{D} & = 0,\\
        \nabla\times\nabla\times \bm{E} & = -\frac{1}{c^2}\frac{\partial^2 \bm{D}}{\partial t^2},
    \end{split}
\end{equation}
where the displacement field $\bm{D}$ and the electric field $\bm{E}$ are related through 
\begin{equation}
    \bm{D} = \varepsilon \bm{E},
\end{equation}
for the fluctuating dielectric field $\varepsilon$.\\
An electromagnetic wave is affected by density changes of the medium it passes through. More specifically, the dielectric $\varepsilon$ and the electric field $\bm{E}$ can be expanded as 
\begin{equation}\label{expansionepsE}
    \begin{split}
        \varepsilon(\bm{x},t) & = \varepsilon_0 + \varepsilon_1 (\bm{x},t) + ...,\\
        \bm{E}(\bm{x},t) & = \bm{E}_{\rm inc}(\bm{x},t) + \bm{E}_1 (\bm{x},t) + ...,
    \end{split}
\end{equation}
where
\begin{equation}\label{defeps1}
    \varepsilon_1(\bm{x},t) = \left.\frac{ \partial \varepsilon}{ \partial \rho}\right|_{\rho = \rho_0} \rho (\bm{x},t),
\end{equation}
for the equilibrium density $\rho_0$ and the deviation from equilibrium $\rho$. Plugging expansion \ref{expansionepsE} into Maxwell's equation \ref{Maxwell} allows us to recover the contributions order by order. In particular, at leading order, we find that $\bm{E}_1$ solves the Helmholtz equation,
\begin{equation}\label{firstorder}
    \nabla^2\bm{D}_1 -\frac{\varepsilon_0}{c^2}\frac{\partial^2 \bm{D}_1}{\partial t^2} = - \nabla \times \nabla \times (\varepsilon_1 \bm{E}_{\rm inc}),
\end{equation}
where $\bm{D}_1 =\varepsilon_0 \bm{E}_1 +\varepsilon_1 \bm{E}_0$. \eqref{firstorder} can be solved assuming the Born approximation (assuming that $\varepsilon_1$ and $ \rho$ vanish outside a domain whose radius is very large compare to its volume, see \cite{landau2013electrodynamics} for details), giving
\begin{equation}\label{defE1}
    \tilde{\bm{E}}_1(\bm{x}, \omega_f) = - \frac{\hat{\bm{x}}\times\hat{\bm{x}}\times \bm{\xi}_0}{\sqrt{2/\pi}c^2}\frac{\omega_f^2 e^{\ri k_f x}}{x} \tilde{\varepsilon}_1(\bm{k}_f-\bm{k}_{\rm inc},\omega_f-\omega_{\rm inc}),
\end{equation}
where $\tilde{\bm{E}}_1$ is the temporal Fourier transform of $\bm{E}_1$, $k_f = \sqrt{\varepsilon_0}\omega_f/c$, $\bm{k}_f = k_f \bm{x}$ is the unit vector along the $\bm{x}$-direction, $x= |\bm{x}|$, $\hat{\bm{x}} = \bm{x}/x$ and $\tilde{\varepsilon}_1(\bm{k},\omega)$ is the spatio-temporal Fourier transform of of $\varepsilon_1$. The field $\bm{E}_1$ defines the leading-order approximation of the electric field of the scattered electromagnetic wave. \\
The Rayleigh--Brillouin spectra are given as the intensity of the scattered electromagnetic wave $\bm{E}_1$,
\begin{equation}\label{IfromE1}
    I(\bm{x},\omega_f) = \langle \bm{E}_1^2\rangle(\bm{x},\omega_f) = \frac{\sqrt{2\pi}}{T} |\bm{E}_1(\bm{x},\omega_f)|^2,
\end{equation}
where $\bm{E}_1$ os assumed to be a periodic function in time. Plugging \eqref{defE1} into \eqref{IfromE1} then gives
\begin{equation}
    |\bm{E}_1(\bm{x},\omega_f)|^2 = \frac{|\bm{\xi}_0|^3\sin(\psi) \omega_f^4}{2c^4 x^2/\pi}|\tilde{\varepsilon}(\bm{k}_f-\bm{k}_{\rm inc},\omega_f-\omega_{\rm inc})|^2,
\end{equation}
where $\psi$ is the angle between $\bm{x}$ and $\bm{x}_0$. By definition \eqref{defeps1}, we have that
\begin{equation}
    |\tilde{\varepsilon}(\bm{k},\omega)|^2 = \left(\frac{\partial \varepsilon}{\partial \rho}\right)^2|\tilde{\rho}(\bm{k},\omega)|^2, 
\end{equation}
where $\tilde{\rho}(\bm{k},\omega)$ is the spatio-temporal Fourier transform of $\rho$. \\
Combining the above equations then leads to the desired relation between the intensity and the density-fluctuations
\begin{equation}
    I(\bm{x},\omega_f) = \frac{V}{(2\pi)^{3/2}}\frac{|\bm{\xi}_0|^2\sin(\psi)\omega_f^4}{2c^4 x^2/\pi}\left(\frac{\partial \varepsilon}{\partial \rho}\right)^2 \langle  \rho ^2\rangle(\bm{k}_f-\bm{k}_{\rm inc},\omega_f-\omega_{\rm inc}).
\end{equation}

\subsection*{Density Fluctuations derived from Fluctuation Spectra}
\label{densityspectra}

In this section, we calculate the density fluctuation spectra for the moment system obtained from \eqref{maineq}, see \cite{lifshitz2013statistical} for more details. To this end, let us first recall the general form of the moment system and how higher-order fluxes enter the dynamics. Taking moments for the density, velocity and the temperature in \eqref{maineq} leads to the general moment system in index form

\begin{equation}\label{moment1}
\begin{split}
       \frac{\partial \rho}{\partial t} + \frac{\partial u_i}{\partial x_i} &= 0 \\
       \frac{\partial u_i}{\partial t} + \frac{\partial  T }{\partial x_i}+ \frac{\partial  \rho }{\partial x_i}  &=  - \frac{\partial \sigma_{ij}}{\partial x_j} \\
\frac{3}{2} \frac{\partial  T  }{\partial t} + \frac{\partial   u_i}{\partial x_i}  &=  -\frac{\partial q_i}{\partial x_i} 
\end{split}
\end{equation}
where 

\begin{equation}
    \bm{\sigma} = \frac{1}{3}\int  ( 3 (\bm{v}\otimes \bm{v}) - |\bm{v}|^2\text{Id}) f(\bm{v})\,d\bm{v},\quad \bm{q} = \frac{1}{2}\int (|\bm{v}|^2-5)\bm{v} f(\bm{v})\, d\bm{v},
\end{equation}
are the stress tensor and the heat flux and Einstein's summation convention has been employed for the velocity field $\bm{u} = (u_i)$ and $\bm{\sigma} = (\sigma_{ij})$.

System \eqref{moment1} is not closed with respect to the macroscopic variables density, velocity and temperature as the stress tensor and the heat flux enter as forcing terms. To close system \eqref{moment1}, we need to impose a constitutive law for the higher-order moments in terms of lower-order moments $\bm{\sigma} = \bm{\sigma}(\rho, \bm{u}, T)$ and $\bm{q} =  \bm{q}(\rho, \bm{u}, T)$. \\
We decompose the velocity field and the heat flux in its transversal and longitudinal part,
\begin{equation}
\bm{u} = \bm{u}_{\perp} + \bm{u}_{\parallel},\quad \bm{q} = \bm{q}_{\perp} + \bm{q}_{\parallel},
\end{equation}
such that $\nabla\cdot \bm{u}_{\perp} = 0$ and $\nabla\cdot \bm{q}_{\perp} = 0$. Similarly, we decompose the stress tensor as the sum of two symmetric tensors 
\begin{equation}
    \bm{\sigma} = \bm{\sigma}_H + \bm{\sigma}_s,
\end{equation}
where $\bm{\sigma}_H$ is the Hessian of a scalar function and $\bm{\sigma}_{s}$ is the sinusoidal part of $\bm{\sigma}$. 

We start with the general linear system of conservation laws in \eqref{moment1}. Taking correlations of \eqref{moment1} with the density and using \eqref{commder} leads to

\begin{equation}\label{corrdensity}
\begin{split}
       \frac{\partial \left<  \rho^2 \right>}{\partial t} + \frac{\partial \left< \rho u_{\parallel i} \right> }{\partial x_i} &= 0 \\
       \frac{\partial \left<  \rho u_{\parallel i} \right>}{\partial t} + \frac{\partial \left< \rho  T\right> }{\partial x_i}+ \frac{\partial \left< \rho^2 \right> }{\partial x_i} + \frac{\partial   \left<  \rho \sigma_{H,ij} \right> }{\partial x_j} &= 0\\
   \frac{\partial \left< \rho u_{\perp j} \right>}{\partial t} +\frac{\partial \left< \rho \sigma_{s,ij} \right> }{\partial x_l} &= 0\\
 \frac{3}{2} \frac{\partial \left< \rho T \right>  }{\partial t}+  \frac{\partial \left< \rho u_{\parallel i} \right>}{\partial x_i} + \frac{\partial \left< \rho  q_{\parallel j} \right>}{\partial x_j} &= 0,
\end{split}
\end{equation}
which will be the basis for our further analysis.\\
We assume that the initial correlations of density with the other macroscopic observables vanish due to statistical independence of the fields. The initial condition for the two-point correlation of density is given by
\begin{equation}\label{initial}
\langle \rho^2 \rangle|_{t=0} = \frac{m N_{\rm eff}}{\rho_0 \Delta x ^3} \delta(\bm{x}),
\end{equation}
where $m$ is the molecular mass, $N_{\rm eff}$ is the effective number of molecules per particle in the DSMC simulation, $\rho_0$ is the equilibrium gas density and $\Delta x$ is the reference length scale used for the non-dimensionalization, see \cite{lifshitz2013statistical} for details.\\
Taking a one-sided temporal and a full spatial Fourier transform of \eqref{corrdensity}, see Appendix \ref{AppFourier}, and using the initial condition \ref{initial} leads to the following linear system in frequency space:
\begin{small}
\begin{equation}\label{spectralequation}
\begin{split}
       \ri \omega \left<  \rho^2 \right>^{+}_{\omega,\bm{k}} + \ri k_i \left< \rho u_{\parallel i} \right>^{+}_{\omega,\bm{k}}   =  \frac{m N_{\rm eff}}{ (2\pi)^2 \rho_0 \Delta x^3  } & \\
       \ri \omega \left< \rho u_{\parallel i} \right>^{+}_{\omega,\bm{k}} + \ri k_i \left< \rho  T\right>^{+}_{\omega,\bm{k}}+ \ri k_i \left< \rho^2 \right>^{+}_{\omega,\bm{k}} + \ri k_j \left<  \rho \sigma_{H,ij} \right>^{+}_{\omega,\bm{k}} &= 0\\
   \ri\omega \left<  \rho u_{\perp i} \right>^{+}_{\omega,\bm{k}} + \ri k_l \left< \rho \sigma_{s,il} \right>^{+}_{\omega,\bm{k}}  &= 0\\
 \ri \omega\frac{3}{2} \left< \rho T \right>^{+}_{\omega,\bm{k}} + \ri k_i \left< \rho u_{\parallel i} \right>^{+}_{\omega,\bm{k}} + \ri k_j \left< \rho  q_{\parallel j} \right>^{+}_{\omega,\bm{k}} &= 0
\end{split}
\end{equation}
\end{small}
Since the longitudinal part is aligned with the wave-vector,
\begin{equation}
\left< \rho u_{\parallel i} \right>^{+}_{\omega,\bm{k}} = \left< \rho u_{\parallel} \right>^{+}_{\omega, k } \hat{k}_i
\end{equation}
transversal and longitudinal part of system \eqref{spectralequation} become independent and the longitudinal part takes the simpler form
\begin{small}
\begin{equation}
\begin{split}
       \ri \omega \left< \rho^2 \right>^{+}_{\omega,\bm{k}} + \ri k \left< \rho u_{\parallel} \right>^{+}_{\omega, k }   =  \frac{m N_{\rm eff}}{(2\pi)^2 \rho_0 \Delta x^3  } &  \\
       \ri \omega  \left< \rho u_{\parallel} \right>^{+}_{\omega, k } + \ri k \left< \rho^2 \right>^{+}_{\omega,\bm{k}} + \ri k \left< \rho T\right>^{+}_{\omega,\bm{k}} + \ri \frac{k_i k_j}{k} \left< \rho \sigma_{H,ij} \right>^{+}_{\omega,\bm{k}} &= 0\\
 \ri\omega \left< \rho T\right>^{+}_{\omega,\bm{k}} + \ri \frac{2}{3} k \left< \rho u_{\parallel} \right>^{+}_{\omega, k }  + \ri \frac{2}{3}  k \left< \rho q_{\parallel} \right>^{+}_{\omega,k}  &= 0. 
\end{split}
\end{equation}
\end{small}
Substituting the constitutive relations gives
\begin{equation} \label{solve spectra}
\begin{split}
       \ri \omega \left< \rho^2 \right>^{+}_{\omega,\bm{k}} + \ri k \left< \rho u_{\parallel} \right>^{+}_{\omega, k}   =  \frac{m N_{\rm eff}}{ (2\pi)^2 \rho_0 \Delta x^3  } & \\
       \ri \omega  \left< \rho u_{\parallel} \right>^{+}_{\omega, k } - \ri \tau_{1} \left< \rho^2 \right>^{+}_{\omega,\bm{k}}  - \tau_{2}\left< \rho u_{\parallel} \right>^{+}_{\omega, k }  - \ri  \tau_{3}\left< \rho T\right>^{+}_{\omega,\bm{k}}     &= 0\\
\ri \omega \left< \rho T \right>^{+}_{\omega,\bm{k}}  -    \tau_{4}\left< \rho^2 \right>^{+}_{\omega,\bm{k}} -  \ri \tau_{5} \left< \rho u_{\parallel} \right>^{+}_{\omega, k }  - \tau_{6}  \left< \rho T \right>^{+}_{\omega,\bm{k}}  &= 0
\end{split}
\end{equation}
Finally, solving for the Fourier transform of the two-point correlation function thus gives the following relation for the fluctuation spectra:

\begin{equation}\label{densitytransportrelation}
 \left<  {\rho}^2 \right>^{+}_{\omega,\bm{k}} = -\frac{\ri \delta_0 (\tau_{3} \tau_{5}+(\tau_{2}-\ri \omega ) (\tau_{6}-\ri \omega ))}{ (\tau_6-\ri\omega)(\omega \tau_2-\ri\omega^2-\ri k \tau_1)+\tau_3(\ri k \tau_4 +\omega \tau_5)},
\end{equation}
for the constant
\begin{equation}
    \delta_0 = \frac{m N_{\rm eff}}{4\pi^2 \rho_0\Delta x^3}. 
\end{equation}

An analogous calculation as outlined above can be carried out for longitudinal velocity correlations. 

\section{DSMC Simulations}
\label{AppDSMC}

In this appendix, we give detailed information on the numerical implementation of the DSMC computation. We use the following non-dimensionalization of physical quantities:

\begin{equation}\label{Fluctuation of the Boltzmann Equation: dimensionless variables}
	\begin{split} 
    \Delta x &= \frac{10 \mu}{\rho_0 } \sqrt{\frac{m}{k_B T_0}}, \quad 
     \Delta t =  \Delta x \sqrt{\frac{m }{k_B T_0}}\\
		\tilde{t} &= \frac{t}{\Delta t},\quad 
		\tilde{x} = \frac{x}{\Delta x},\quad  
		\tilde{\bm{u}} = \frac{\Delta t }{\Delta x} \bm{u},\quad 
		\tilde{\rho} = \frac{\rho}{\rho_0},\quad 
		\tilde{T} = \frac{T}{T_0} \\
	\end{split}
\end{equation}
where the reference values of the dynamic viscosity $\mu$, the equilibrium density $\rho_0$, the equilibrium temperature$T_0$ and the molecular mass $m$ are specified in Table.\ref{AP_tab:table2}. The global Knudsen number used in the DSMC simulation is defined as
\begin{equation} \label{Kn_sys}
    \text{Kn} = \frac{\mu}{\rho \Delta x} \sqrt{\frac{m}{k_B T}},
\end{equation}
where $\Delta x$ is the reference length used to non-dimensionalize the spatial coordinates in numerical computations, $\mu$ is the dynamic viscosity, $\rho$ is the reference density, $m$ is the molecular mass, $k_B$ is Boltzmann's constant and $T$ is the reference temperature.
\begin{table}
\caption{\label{AP_tab:table2}
Parameters used in 
DSMC simulation and moment advection computation}
\resizebox{0.5\textwidth}{!}{%
\begin{tabular}{l|l|l|l}
Domain size & $1.125 \times 0.25^2 \,\mathrm{m}^3$  & Collision Model & VHS\\
Power law $\gamma$ & 1 & Diameter & $4.17\times10^{-10}\, \mathrm{m}$\\
Number of Cells & $225 \times 50 \times 50$ & Mean Free Path & $1.35\times10^{-2}\,\mathrm{m}$ \\
Number of Particles & $7031250$ & Mean Free Time & $3.40\times10^{-5}\,\mathrm{s}$ \\
Density $\rho$ & $6.63\times10^{-6}\,\mathrm{kg}/\mathrm{m}^3$ & Temperature $T$ & $300\,\mathrm{K}$ \\
Molecule Mass $m$ & $6.63\times10^{-26} \, \mathrm{kg}$  
& Sound Speed & $322.68\, \mathrm{m}/\mathrm{s}$ \\
Heat Conduction & $0.022\, \mathrm{W}/(\mathrm{m K})$ & Viscosity $\mu$  &$2.82\times10^{-5}\, \mathrm{Pa}\,\mathrm{s}$ \\
Time step size & $2\times10^{-5}\, \mathrm{s}$ & Cell width & $5\times10^{-3}\, \mathrm{m}$\\
Subcell & 1 & Simulation time& $12.10\, \mathrm{s}$  \\
\end{tabular}
}
\end{table}

We use a re-implemented parallelized version the DSMC3 program by Bird \cite{Bird1994MolecularGDF} to simulate the fluctuation of a three-dimensional homogeneous gas. The domain of the simulation is a rectangular cuboid with periodic boundary condition. The cuboid has a spatial span of $1.125m$ in the x-direction and is divided uniformly into $225$ cells, while its spatial span in the y- and z-directions is $0.25m$, where it is divided uniformly into $50$ cells. Each cell contains one sub-cell utilized in determining collision pairs in the DSMC computation. The initial condition of our DSMC computation uses particle velocities sampled from a Maxwell distribution with $T=300K$ and zero mean velocity. The particle position is uniformly distributed in each cell. More details about the properties of the gas are shown in Table \ref{AP_tab:table2} using SI units. 

The merit of the DSMC calculation is that no driven physical conditions are required for simulating fluctuations, since the DSMC method uses Monte Carlo samples to mimic the real gas molecules. Statistical quantities computed from Monte Carlo samples therefore naturally fluctuate in the same way as the real gas except for an enlarged fluctuation amplitude. Specifically, if one sample in the DSMC simulation represents $N_{\rm eff}$ real gas molecules, the variances of fluctuations in statistical quantities computed from the DSMC simulation are $N_{\rm eff}$ times larger than those of a real gas. In our DSMC computation we have $7031250$ simulation particle samples representing gases of number density $10^{20}m^{-3}$ and each sample particles represents $N_{\rm eff}=3.75\times 10^{15}$ real gas molecules.


The molecular model is crucial in DSMC calculations. It describes how two molecules collide with each other and determines the viscosity of the gas. The molecular model gives the relation between two characteristic quantities of a classical binary collision problem: the impact parameter $b$ and scattering angle $\theta$. A typical molecular model used in DSMC is the variable hard/soft sphere model \cite{Bird1994MolecularGDF}, 

\begin{equation} \label{VHS model: The deflection angle}
	\theta = 2 \arccos\left(\left(\frac{b}{d}\right)^{\frac{1}{\alpha}}\right)
\end{equation}
where $d$ is the effective diameter of the gas molecules and $\alpha$ is a parameter mainly effecting the diffusion coefficient. The diffusion parameter describes mass diffusion between components of gas mixtures and is irrelevant in our single species case. Therefore we use the default value $\alpha=1$ corresponding to the variable hard sphere model (VHS). The effective diameter $d$ varies with the relative velocity between colliding molecules
\begin{equation} \label{DSMC: the effective diameter, calculation, final}
	d = d_{\rm ref}\left(\frac{(2k_B T_{\rm ref}/(\frac{1}{2}m v_r^2))^{\gamma-1/2}}{\Gamma(5/2-\gamma)}\right)^{1/2}
\end{equation}
where $m$ is the mass of a gas molecule, $v_r$ is the relative velocity between the two colliding molecules, $\Gamma$ represents the Gamma function, $T_{\rm ref}$ is the reference temperature, $d_{\rm ref}$ is the reference molecule diameter, and $\gamma$ is a parameter that determines how viscosity coefficient changes with respect to temperature. In our computation we use the default values $m=6.63\times10^{-26} kg$, $T_{\rm ref}=273K$, and $d_{ref}=4.17 \times 10^{-10}m$. Note that \eqref{DSMC: the effective diameter, calculation, final} differs from the equation in \cite{Bird1994MolecularGDF} since the authors use the reduced mass $m_r=\frac{1}{2}m$ instead of molecule mass $m$ in our case. The parameter $\gamma$ in \eqref{DSMC: the effective diameter, calculation, final} determines the power law between the viscosity coefficient $\mu$ and the temperature $T$ in the form $\mu \propto T^\gamma$. The choice $\gamma=1$ corresponds to Maxwell molecules. 


We compute the viscosity coefficient and heat conduction coefficient of our DSMC simulated gas using the Chapman-Enskog theory
\begin{equation} \label{DSMC viscosity coef}
\begin{split}
    	\mu &= \frac{5(\alpha+1)(\alpha+2)(\pi m k_B)^{1/2}(4k_B/m)^{\gamma-1/2} T^\gamma}{16\alpha\Gamma(9/2-\gamma)\sigma_{T,{\rm ref}}v_{r,{\rm ref}}^{2\gamma-1}},\\
	\kappa &= \frac{15 k_B}{4 m}\mu,
\end{split}
\end{equation}
for the reference total cross section $\sigma_{T,{\rm ref}} =\pi d_{\rm ref}^2 $ and the reference velocity
\begin{equation}
    v_{r,{\rm ref}}  = \sqrt{\frac{4k_B T_{\rm ref}}{m\Gamma(5/2-\gamma)^{\frac{1}{\gamma-1/2}}}}.
\end{equation}

To ensure the resolution at relatively large Knudsen numbers in our DSMC computation, we set the cell width to be five times smaller than the mean free path of the gas, while the time step is set to be ten times smaller than the mean free time of the gas.  We compute the mean free path and the mean free time from the collision rate per gas molecule according to 
\begin{equation} \label{DSMC: the collision frequency, general, deff, explicit, numerical}
	f = 4 n \sqrt{\pi} d_{\rm ref}^2  \left(\frac{ T}{T_{\rm ref}}\right)^{\frac{1}{2}-\gamma}\left(\frac{k_B T}{m}\right)^{\frac{1}{2}},
\end{equation}
where $n$ is the number density, related to the mass density and the molecular mass by $n=\rho/m$.

Our computation is independent of the mean flow component since we are simulating a homogeneous gas with homogeneous initial conditions. The finite simulation domain in our DSMC computation, however, may introduce deviation in the spectra from the theoretically predicted results. To eliminate this finite domain effect, we use a domain length much larger than the mean free path of the gas in the x-direction, along which the fluctuation spectra is measured, to ensure consistency.  A snap shot of the simulation is saved every two time steps. Then, the macroscopic quantities for each cell are calculated by averaging the corresponding quantities of the particles in each cell. The density fluctuation spectra used to train the neural net are computed via the discrete Fourier transformation of the density profile according to  \eqref{WienerKinchin}.

\section{The Shakhov Model: Explicit Spectral Closure and Monte Carlo Simulations}
\label{AppShakhov}

In this appendix, we recall the linearized Shakhov model in dimensional and non-dimensional form. We discuss its numerical solution by a pseudo-spectral Monte Carlo method. Finally, we recall the spectral properties of the linearized Shakhov model and derive the exact spectral closure. 

\subsection*{The linear Shakhov Equation} 
The Shakhov equation is a quasi-equilibrium kinetic equation, whose collision operator only depends on density, velocity, temperature and heat flux \cite{shakhov1968generalization}. It is a generalization of the BGK collision model and allows for different Prandtl numbers, in particular the physically relevant case $\mathrm{Pr}=2/3$. 

In dimensional form, the linearized Shakhov equation is given by
\begin{equation}
\label{Shakhov}
\begin{split}
\frac{\partial f}{\partial t} 
&+ \bm{v}\cdot\bm{\nabla}f 
= - \frac{(1 - \text{Pr}) m (\bm{v} \cdot \bm{q})}{\tau n_0 k_B^2 T_0^2}
   \left(1 - \frac{m\bm{v}^2}{5 k_B T_0 }\right) \notag \\[6pt]
&+ \frac{n}{\tau n_0}
+ \frac{m \bm{v} \cdot \bm{u}}{\tau k_B T_0} 
+ \left(\frac{m \bm{v}^2}{2 k_B T_0} - \frac{3}{2}\right)
   \frac{T}{\tau T_0} \notag - \frac{f}{\tau},
   \end{split}
\end{equation}
for the unknown distribution function $f$, the global relaxation time $\tau$, the molecular mass $m$, the Prandtl number Pr, the reference temperature $T_0$, the reference number density $n_0$ and Boltzmann's constant $k_B$. The macroscopic variables in dimensional form, linearized around the global equilibrium distribution 
\begin{equation}\label{globalMax}
   \phi(\bm{v}) = n_0\left(\frac{2\pi k_B T_0}{m}\right)^{-\frac{3}{2}} e^{-\frac{m|\bm{v}|^2}{2k_B T_0}} ,
\end{equation}
are given by
\begin{equation} 
\label{Fluctuation of the Boltzmann Equation: fields operator_swapped}
\begin{split}
     n 
	 & = \int  f\,\phi\, d^3 \bm{v},\\
     \bm{u} 
	& = \frac{1}{n_0} \int \bm{v}\,  f\,\phi\, d^3 \bm{v},\\
     T  
   & = \frac{m}{3\, n_0\, k_B} 
     \int \!\, |\bm{v}|^2\, f\, \phi\, d^3 \bm{v}
   \;-\;  T_0 \,\frac{ n}{n_0},\\
   \bm{q}
	& = \frac{m}{2}\int \,\bm{v}\,|\bm{v}|^2\, f\, \phi\, d^3 \bm{v}
	  \;-\;\tfrac{5}{2}\,\bm{u}\, n_0\,k_B\,T_0.
\end{split}
\end{equation}
\eqref{Shakhov} is derived from the full non-linear Shakhov model by assuming a solution of the form $F = (1+f)\phi$, where $\phi$ is the global Maxwellian (\ref{globalMax}). 

Applying the one-sided temporal Fourier transform as defined in \eqref{defonesidedFourier} and the full spatial Fourier transform as defined in \eqref{defFouriertransform} to \eqref{Shakhov}, we obtain
\begin{align}
\label{Fluctuation of the Boltzmann Equation: The linearized Boltzmann Equation noise FFT_swapped}
\ri \omega \hat{f}^{+} 
&+ \ri (\bm{v} \cdot \bm{k}) \hat{f}^{+} \notag = - \frac{(1 - \mathrm{Pr}) m}{\tau  n_0 k_B^2 T_0^2} 
   \left(1 - \frac{m\bm{v}^2}{5 k_B T_0 }\right) 
   \bm{v} \cdot \hat{\bm{q}}^{+} \notag \\[6pt]
&\quad + \frac{ \hat{n}^{+}}{\tau n_0} 
   + \frac{m \bm{v} \cdot \hat{\bm{u}}^{+}}{\tau k_B T_0} \notag  + \left(\frac{m \bm{v}^2}{2 k_B T_0} - \frac{3}{2}\right) 
   \frac{\hat{T}^{+}}{\tau T_0} \notag \\[6pt]
&\quad - \frac{\hat{f}^{+}}{\tau} 
   + \frac{\hat{f}(\bm{v}, \bm{k}, t=0)}{\sqrt{2\pi}}.
\end{align}
The non-dimensionalization 
in \eqref{Fluctuation of the Boltzmann Equation: dimensionless variables} brings equation \eqref{Fluctuation of the Boltzmann Equation: The linearized Boltzmann Equation noise FFT_swapped} into the form
\begin{align}
\label{Fluctuation of the Boltzmann Equation: The linearized Boltzmann Equation noise FFT non-dim_swapped}
\ri \,\tilde{\omega}\,\tilde{f}^{+} 
&+ \ri\,(\tilde{\bm{v}} \cdot \tilde{\bm{k}})\,\tilde{f}^{+} = -\,\frac{(1 - \mathrm{Pr})}{\mathrm{Kn}} 
    \left(1 - \frac{\tilde{\bm{v}}^2}{5}\right) 
    \tilde{\bm{v}} \cdot \tilde{\bm{q}}^{+} \notag \\[6pt]
&\quad 
    +\,\frac{\tilde{n}^{+}}{\mathrm{Kn}} 
    +\,\frac{\tilde{\bm{v}} \cdot \tilde{\bm{u}}^{+}}{\mathrm{Kn}} \notag +\,\left(\frac{\tilde{\bm{v}}^2}{2} - \frac{3}{2}\right)\,
    \frac{\tilde{T}^{+}}{\mathrm{Kn}} \notag \\[6pt]
&\quad -\,\frac{\tilde{f}^{+}}{\mathrm{Kn}} 
 +\,\frac{\tilde{f}(\tilde{\bm{v}}, \tilde{\bm{k}}, \tilde{t}=0)}{\sqrt{2\pi}},
\end{align}
for the non-dimensional moments 
\begin{equation} \label{Fluctuation of the Boltzmann Equation: Density operator one side nodim_swapped}
\begin{split}
	\tilde{n}^{+}
	\; & =\;\int \tilde{f}^{+} \exp\Bigl(-\tfrac{\tilde{\bm{v}}^2}{2}\Bigr)\,d^3\tilde{\bm{v}},\\
    \tilde{\bm{u}}^{+}
	\; & =\;\frac{1}{(2\pi)^{3/2}}
	     \int \tilde{\bm{v}}\tilde{f}^{+}\,
	     \exp\Bigl(-\tfrac{\tilde{\bm{v}}^2}{2}\Bigr)\,
	     \,d^3\tilde{\bm{v}},\\
         \tilde{T}^{+}
	& \;=\;\frac{1}{(2\pi)^{3/2}}\,
	     \frac{1}{3}\,\int 
	     |\tilde{\bm{v}}|^2\tilde{f}^{+}\,
	     \exp\Bigl(-\tfrac{|\tilde{\bm{v}}|^2}{2}\Bigr)\,\,
	     d^3\tilde{\bm{v}}
	     \;-\; \tilde{n}^{+}\\
         \tilde{q}^{+}_i
	\; & =\;\frac{1}{(2\pi)^{3/2}}
	     \int \frac{1}{2}\,\tilde{v}_i\,
	           |\tilde{\bm{v}}|^2 \tilde{f}^{+}\,
	           \exp\Bigl(-\tfrac{|\tilde{\bm{v}}|^2}{2}\Bigr)\,
	           \,d^3\tilde{\bm{v}}
	     \;-\;\frac{5}{2}\,\tilde{u}_i^{+}.
    \end{split}
\end{equation}

\subsection*{Monte Carlo Simulations for the Shakhov Model}

In the following, we provide details for the numerical solution of the Shakhov model.

To compute the density fluctuations, we impose the initial condition
\begin{equation}
	\tilde{f}_0(\tilde{\bm{v}}, \tilde{\bm{k}}) \;=\;\frac{1}{(2\pi)^{3/2}}\,\frac{m\,N_{\mathrm{eff}}}{\rho_0\,(\Delta x)^3},
\end{equation}
which corresponds to the initial density 
\begin{equation}
	\tilde{\rho}_0(\bm{x})
\;=\;\frac{m\,N_{\mathrm{eff}}}{\rho_0}\,\delta(\bm{x}),
\end{equation}
for the Dirac delta distribution $\delta$. Thanks to a rotational symmetry, we know that if $\tilde{f}^{+}$ is a solution with non-dimensional wave vector $\tilde{\bm{k}}$, so is the rotated version of that solution with wave vector $\bm{O}\tilde{\bm{k}}$ for any rotation matrix $\bm{O}$. Consequently, we may restrict the solution procedure to wave vectors of the form $\tilde{\bm{k}} = (\tilde{k}, 0, 0)$ without loss of generality.

To solve \eqref{Fluctuation of the Boltzmann Equation: The linearized Boltzmann Equation noise FFT non-dim_swapped} numerically, we first split it into real and imaginary part and then use Anderson Acceleration \cite{10.1145/321296.321305}, i.e., an iterative fixed-point approach. 
In each iteration, the real and imaginary parts of $\tilde{f}^{+}$ are updated by evaluating the Gaussian integrals  for $\tilde{n}^{+}, \tilde{\bm{u}}^{+},  \tilde{T}^{+}$  and $\tilde{\bm{q}}^{+}$ by a quasi-Monte Carlo method with $2^{14}$ sample particles, until convergence is achieved with residual less than $10^{-8}$.

We emphasize that the Monte Carlo integration scheme performs considerably better compared to integration with Hermite polynomials, i.e., integration by projection onto an orthogonal basis. Indeed, projection onto Hermite polynomials can lead to nonphysical spurious peaks in the computed spectra.


\subsection*{The Spectral Closure of the Shakhov Model}

In this appendix, we describe the spectral closure for the Shakhov model in detail. As discussed before, the theory of spectral closure was derived for general linear Boltzmann-type equations in \cite{PhysRevE.110.055105} and carried out for the linear BGK model in \cite{kogelbauerBGKspectral2,PhysRevE.110.055105}. We emphasize that this is the first explicit calculation of the spectral closure for the Shakhov model.\\
Our starting point for obtaining the exact transport coefficients is the non-dimensional linearized Shakhov model in \eqref{Shakhov}, rewritten in the form
\begin{equation}\label{eqmainlinear}
    \frac{\partial f}{\partial t} = \mathcal{L}f,
\end{equation}
for an unknown scalar distribution function $f$ and the linear operator
\begin{equation}\label{defL}
\mathcal{L}=-\bm{v}\cdot\nabla_{\bm{x}}-\frac{1}{\tau}(1-\mathbb{B}_{8,\rm Pr}),
\end{equation}
consisting of the free-transport part and the linear collision part 
\begin{equation}
\mathbb{B}_{8,\rm Pr}=\mathbb{P}_5+(1-\rm Pr)\mathbb{P}_8,
\end{equation}
where Pr is the Prandtl number. The projection operators $\mathbb{P}_5$ and $\mathbb{P}_8$ are given by
\begin{equation}
\mathbb{P}_5f = \sum_{n=0}^4 \langle f,e_n\rangle e_n,\qquad  \mathbb{P}_8f = \sum_{n=5}^7 \langle f,e_n\rangle e_n,
\end{equation}
where in the inner product $\langle . , .\rangle$ is relative to the non-dimensional Maxwellian in \eqref{globalMax}, for the following set of moment functions.
\begin{align}\label{basis}
e_0(\bm{v}) &= 1,\qquad e_4(\bm{v}) = \frac{|\bm{v}|^2-3}{\sqrt{6}},\\
e_1(\bm{v}) &= v_1,\qquad e_5(\bm{v}) = v_1\frac{|\bm{v}|^2-5}{\sqrt{10}},\\
e_2(\bm{v}) &= v_1, \qquad e_6(\bm{v})  = v_2\frac{|\bm{v}|^2-5}{\sqrt{10}},\\
e_3(\bm{v}) &= v_3, \qquad e_7(\bm{v}) = v_3\frac{|\bm{v}|^2-5}{\sqrt{10}}.
\end{align}
The basis functions \eqref{basis} satisfy the orthonormality condition
\begin{equation}
\langle e_n, e_m \rangle_{\bm{v}} = \delta_{nm}, \quad \text{for} \quad  0\leq n,m \leq 7.
\end{equation}
To ease notation, we bundle the eight moment functions \eqref{basis} into a single vector,
\begin{equation}
e=\{e_j\}_{0\leq j\leq 7},
\end{equation}
and define the Prandtl-number-dependent matrix
\begin{equation}
D_{\Pr}=\text{diag}(\text{Id}_{5\times 5},(1-\rm Pr) \text{Id}_{3\times 3}).
\end{equation}
As in the previous section, we apply a spatial Fourier transform to the operator in \eqref{defL} to 
conjugate the operator \eqref{defL} to a wave-number dependent family of linear operators
\begin{equation}\label{defLhat}
\mathcal{L}_{\bm{k}}=-\ri\bm{k}\cdot\bm{v}-\frac{1}{\tau}+\frac{1}{\tau}\mathbb{B}_{8,r}.
\end{equation}

For details of the calculation of the spectrum of the Shakhov model, we refer to \cite{KOGELBAUER2024134014}, including the derivation of a spectral function, the explicit form of the eigenvectors, the discussion of branch merging and the description of the hydrodynamic manifold. We recall that the spectrum of the linear Shakhov equation is given by
\begin{equation}
\sigma(\mathcal{L}) = \left\{-\frac{1}{\tau}+\ri\mathbb{R}\right\}\cup\bigcup_{k<k_{\rm crit}}\bigcup_{N\in \text{Modes}(k,\rm Pr)}\{\lambda_{N}(
k)\},
\end{equation}
where, for $\rm Pr = 2/3$, the set of modes is given by
 \begin{equation}\label{mode1thm}
    \rm Modes = \{\rm s_1,d_1,a_1,a_1*,s_2,d_2\},
 \end{equation}
consisting of the primary and secondary, double degenerated shear modes  $\lambda_{\rm s,1},\lambda_{\rm s, 2}\in\mathbb{R}$, the pairs of complex conjugated primary and secondary acoustic mode $\{\lambda_{\rm a,1}, \lambda_{\rm a,1}^*\}, \{\lambda_{\rm a,2}, \lambda_{\rm a,2}^*\}$ and the real primary and secondary diffusion modes $\lambda_{\rm d,1}, \lambda_{\rm d,2}$. \\
The frequency-dependent eigenfunctions solve the equation
\begin{equation}
-\ri\bm{k}\cdot\bm{v}\hat{f}_n -\frac{1}{\tau}\hat{f}_n+\mathbb{B}_{8,\rm Pr} \hat{f}_n = \lambda_n \hat{f}_n,
\end{equation}
which has the explicit solution
\begin{equation}\label{eigf2ts}
\hat{f}_n = \frac{e\cdot\alpha_n}{\tau\ri\bm{k}\cdot\bm{v}+1+\tau\lambda_n}.
\end{equation}
The coefficient vector $\alpha_n$ is defined as
\begin{equation}\label{alpha}
    \alpha_n = \langle  \hat{f}_n,e\rangle_{\bm{v}},
\end{equation}
and satisfies the equation
\begin{equation}
    \alpha_n \in \text{ker}( D_{\rm Pr}\tilde{G}(z,\bm{k})-\text{Id})_{z=-1-\tau\lambda_n},
\end{equation}
for the full Green's matrix for the Shakhov model
\begin{equation}
\tilde{G}(z,\bm{k})=\int_{\mathbb{R}^3} e(\bm{v})\otimes e(\bm{v})\frac{e^{-\frac{|\bm{v}|^2}{2}}}{\tau\ri\bm{k}\cdot\bm{v}-z} \, d\bm{v}.
\end{equation}
Indeed, defining the  \textit{spectral function} of the Shakhov model,
\begin{equation}\label{spectralfunction}
    \Sigma_{\bm{k},\tau}(\lambda) = \det\Big( D_{{\rm Pr}}\tilde{G}(z,\bm{k})-\text{Id}\Big)_{z=-1-\tau\lambda},
\end{equation}
the discrete spectrum (between the essential spectrum and the imaginary axis) is  given by 
\begin{equation}\label{spec1}
\sigma_{\rm disc}(\mathcal{L}_{\bm{k}})=\left\{\lambda \in\mathbb{C}:  \Sigma_{\bm{k},\tau}(\lambda) =0\right\}.
\end{equation}
We are, however, only interested in the first five entries of an eigenvector of $\mathcal{L}_{\bm{k}}$ in the following, which corresponds to the five-by-five submatrix $n\in \{\rm d,s,a,a*\}$ and the first five entries of the alpha-vectors. To ease notation, we bundle the primary hydrodynamic moments, density, velocity and temperature, into a single vector analogously to  \eqref{defh},
\begin{equation}
        h = (\rho,\bm{u},T). 
\end{equation}

While the BGK equation only models the five primary hydrodynamic branches of the spectrum of the full Boltzmann equation \cite{kogelbauer2024exact}, the Shakhov equation allows to resolve the heat flux as a macroscopic variable as well. For Prandtl number $\rm Pr = 2/3$, however, no branch merging occurs in the Shakhov model \cite{KOGELBAUER2024134014} and the primary hydrodynamic branches emerging from the collision invariants define a slow manifold for all wave numbers up to the critical wave number.

The spectral function of the Shakhov operator takes the explicit form 
\begin{equation}\label{Sigmaexpl}
\begin{split}
\Sigma_{\mathbf{k},\tau}(\lambda) & = \frac{1}{3000(\ri \kappa )^8}\Big[ \Sigma_0(\zeta) +\Sigma_1(\zeta)Z(\zeta )+ \Sigma_2(\zeta) Z(\zeta )^2\Big]^2 \\
& \times \Big[\Sigma_3(\zeta)+\Sigma_4(\zeta)Z(\zeta)+\Sigma_5(\zeta)Z^2(\zeta)\Big]_{\zeta=\ri\frac{\tau \lambda+1}{\kappa}},
   \end{split}
\end{equation}
where 
\begin{equation}\label{defZ}
    Z(\zeta) = \frac{1}{\sqrt{2\pi}} \int_{\mathbb{R}} \frac{e^{-\frac{v^2}{2}}}{v-\zeta}\, dv,
\end{equation}
for any $\zeta\in\mathbb{C}\setminus\mathbb{R}$, is the plasma dispersion function and 
\begin{equation}\label{coefpoly}
    \begin{split}
        \Sigma_0(\zeta) & = 10 \kappa ^2+\zeta  r \left(\ri \zeta ^2 \kappa +\zeta -\ri \kappa \right),\\
        \Sigma_1(\zeta) & = 10 \ri \kappa +r \left(\ri \zeta ^4 \kappa +\zeta ^3-2 \ri \zeta ^2
   \kappa -\zeta +9 \ri \kappa \right),\\
   \Sigma_2(\zeta) &= -8r,\\
   \Sigma_3(\zeta)& = 5 \kappa  \left(\ri \zeta ^3 \kappa ^2+2 \zeta ^2 \kappa +\ri \zeta  \left(5 \kappa ^2-1\right)+6 \left(\kappa ^3+\kappa \right)\right)\\
   &\qquad+r \left(3
   \ri \zeta ^5 \kappa ^3+9 \zeta ^4 \kappa ^2-3 \ri \zeta ^3 \kappa  \left(5 \kappa ^2+3\right)\right)\\
   &\qquad+r\left(-\zeta ^2 \left(43 \kappa ^2+3\right)+2 \ri \zeta 
   \kappa  \left(15 \kappa ^2+23\right)+6 \left(5 \kappa ^2+3\right)\right),\\
   \Sigma_4(\zeta)  &= \ri \left(5 \kappa  \left(\zeta ^4 \kappa ^2-2 \ri \zeta ^3 \kappa +\zeta ^2 \left(4 \kappa ^2-1\right)+11 \kappa ^2+5\right)\right.\\
   &\qquad+r \left(3 \zeta ^6
   \kappa ^3-9 \ri \zeta ^5 \kappa ^2-9 \zeta ^4 \left(2 \kappa ^3+\kappa \right)+\ri \zeta ^3 \left(64 \kappa ^2+3\right)\right.\\
   &\left.\left.+\zeta ^2 \kappa 
   \left(39 \kappa ^2+79\right)-\ri \zeta  \left(23 \kappa ^2+33\right)+16 \kappa \right)\right),\\
   \Sigma_5(\zeta) & = -4 \left(5 \kappa  \left(\zeta ^2 \kappa -\ri \zeta +\kappa \right)\right.\\
   &\qquad\left.+\zeta  r \left(3 \zeta ^3 \kappa ^2-6 i \zeta ^2 \kappa +5 \zeta  \kappa
   ^2-3 \zeta -5 \ri \kappa \right)\right).
    \end{split}
\end{equation}
are polynomials. We refer to \cite{KOGELBAUER2024134014} for details of the derivation.

The first critical wave number among the primary hydrodynamic branches is associated to the diffusion mode, 
\begin{equation}
    k_{\rm crit, diff} \approx 0.9650\frac{1}{\tau},
\end{equation}
obtained as the first wave number for which a zero of $\det G$ merges into the essential spectrum.

\begin{figure*}
    \includegraphics[width= 17.8cm]{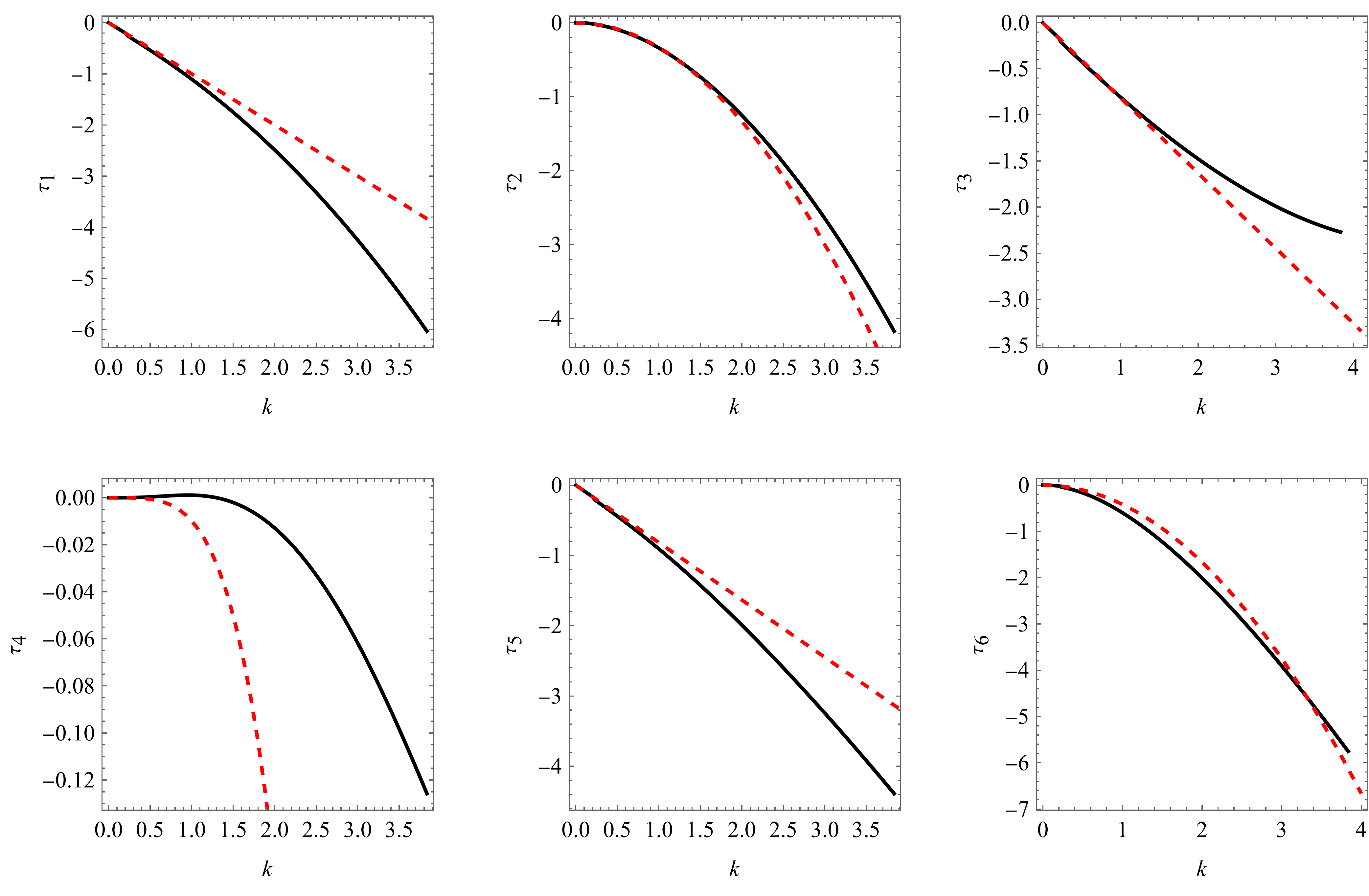}
     \caption{The transport coefficients of the linear Shakhov model in dependence on wave number ($0\leq k\leq k_{\rm crit,diff}$) for $\tau=0.25$ (solid black line) compared to its leading-order approximation at the origin (Burnett/Navier--Stokes/Euler, dashed red line). For small wave numbers, corresponding to small Knudsen numbers, the exact generalized transport coefficients agree well with their corresponding Chapman--Enskog approximation, while for larger wave numbers, the difference to local hydrodynamics is quite pronounced.}
     \label{cplot}
\end{figure*}

The Shakhov transport coefficients \eqref{defTShakhov} are depicted in Figure \ref{cplot}.

\section{Neural Network Architecture and Training Procedure}
\label{AppNeuralNet}

\subsection*{Neural Network Overview}

In this work, we employ a neural network (NN) to infer constitutive laws for a rarefied gas. The NN predicts the wave-number-dependent generalized transport coefficients \(\{\tau_i\}_{1\leq j\leq 6}\), which generalize classical constitutive relations, such as the Chapman--Enskog expansions, and allow us to match the measured density, velocity, and temperature fluctuation spectra.

The neural network framework comprises two components: 
\begin{enumerate}
    \item A parameter-predicting sub-network, \texttt{NetNN}, which maps the input wave number \( k \) to a set of intermediate parameters.
    \item A \texttt{spectral solver}, which uses these parameters to compute the full set of generalized transport coefficients \(\{\tau_i\}_{1\leq j\leq 6}\) and thereafter the fluctuation spectra.
\end{enumerate}

\subsection*{Parameter-Predicting Network (NetNN)}

The network \texttt{NetNN} is defined by a feedforward architecture:
\begin{itemize}
    \item \textbf{Input:} A single scalar input, the wave number \( k \).
    \item \textbf{First Linear Layer:} A fully connected layer \(\text{Linear}(1, 10)\) maps \( k \) to a 10-dimensional latent space, formally written as 
    \[
    h_1 = W_1 k + b_1,
    \]
    where \(W_1 \in \mathbb{R}^{10 \times 1}\), \(b_1 \in \mathbb{R}^{10}\) and \(h_1 \in \mathbb{R}^{10}\).
    \item \textbf{Layer Normalization and Activation:} A layer normalization (LN) and a Gaussian Error Linear Unit (GELU) activation are applied:
    \[
    h_1' = \text{GELU}(\text{LN}(h_1)).
    \]
    \item \textbf{Residual Connection:} A skip-connection linearly transforms the original input \(k\) into the 10-dimensional space and adds it to the activated representation:
    \[
    x_1 = h_1' + W_{s1} k + b_{s1},
    \]
    where \(W_{s1}\) and \(b_{s1}\) are the parameters of this skip connection.
    \item \textbf{Second Linear Layer:} Another linear transformation and normalization step is performed:
    \[
    h_2 = \text{GELU}(\text{LN}(W_2 x_1 + b_2)) + x_1,
    \]
    yielding a final 10-dimensional feature vector. This second residual connection helps stabilizing the training.
    \item \textbf{Output Layer:} Finally, a linear map \(\text{Linear}(10, 2)\) produces a 2-dimensional output vector, which is rescaled for numerical stability:
    \[
    p(k) = 0.1 (W_3 h_2 + b_3).
    \]
    The resulting output
    \begin{equation}
        p(k) = [N_2(k), N_6(k)]
    \end{equation}
    represents intermediate parameters used by the \texttt{spectra\_Net} to determine the effective transport coefficients.
\end{itemize}

\subsection*{Spectral Solver}

The \texttt{spectra Solver} class encodes the physics of the rarefied gas system. Given physical constants, such as the molecular mass \(m\), the density \(\rho\), the temperature \(T\), and the viscosity \(\mu\), along with the output \(p(k)\) from \texttt{NetNN}, it computes the wave-number dependent generalized transport coefficients \(\tau_i = \tau_i(k)\).
In the following, the Knudsen number is defined as in \eqref{Kn_sys}.

The network uses \(N_2(k)\) and \(N_6(k)\) to define the values
\[
\kappa_1 = -\tfrac{4}{3}\text{Kn}\exp(N_2(k)), \quad \kappa_2 = \frac{45}{9}\mu_2\exp(N_6(k)). 
\]
Once the values of the generalized transport coefficients \(\tau_i\) are known, the network evaluates the linearized hydrodynamic spectra for the density \(\rho\), the velocity parallel to the wave vector \(u_{\parallel}\), and the temperature \(T\) at each temporal frequency \(\omega\) and each spatial frequency $k$. This calculation uses the linearized hydrodynamic equations and incorporates the new \(\tau_i\)-based constitutive relations. The resulting spectra $\rho^2(k,\omega), \rho u_{\parallel}(k,\omega)$ and $\rho T(k,\omega)$
are the final outputs that are compared against reference data, e.g., DSMC or Shakhov simulations. One could also compute other spectra like $u_{\parallel}^2(k,\omega)$, $u_{\parallel} T(k,\omega)$, etc.

\subsection*{Training Procedure}

The training procedure  minimizes the mean squared error (MSE) between predicted and reference spectra. Let \(\rho^2_{\rm spec}(k,\omega)\), \(\rho u_{\parallel, \rm spec}(k,\omega)\), and \(\rho T_{\rm spec}(k,\omega)\) be the reference spectra and define the loss function \(\mathcal{L}\) as
\begin{small}
\begin{equation}
L = \left\langle W_{k,\omega}\left[(\rho^2-\rho_{\rm spec})^2 + (\rho u_{\parallel}-\rho u_{\parallel \rm spec})^2 + (\rho T-\rho T_{\rm spec})^2\right]\right\rangle_{k,\omega},
\end{equation}
\end{small}
where \(W_{k,\omega}\) are frequency- and wave-number-dependent weights. These weights normalize the loss by the peak spectral magnitude for each \(k\) and \(\omega\), ensuring stable training and balanced emphasis across scales. The stochastic gradient-based optimizer Adam is used to solve the optimization problem. The learning rate decays linearly over \( \sim1200 \) epochs. Model evaluation on a held-out validation set guides hyper-parameter tuning and early stopping. Saved model states and loss history plots are used to track progress and ensure generalization.

\subsection*{Physical Interpretation}

By adjusting the transport coefficients \(\tau_i\) to fit the reference data, the NN effectively infers how non-equilibrium and rarefaction effects modify linearized hydrodynamic fluctuations beyond classical Navier--Stokes or Chapman--Enskog theory. The trained model thus provides a flexible, data-driven mapping from any wave number \(k\) to the effective transport parameters. This mapping yields a physically interpretable correction to traditional constitutive relations, enabling improved predictions of fluctuation spectra in rarefied flow regimes.

\section{Detailed Neural Network Parameterization, Input Scaling, and Training Weights}
\label{AppLearning}

\subsection*{Computation of \(\tau\)-Curves from NetNN Outputs}

The neural network (\texttt{NetNN}) takes a wave number input \( k \) and produces two dimensionless parameters, \( N_2(k) \) and \( N_6(k) \), which serve as nonlinear corrections to the generalized transport coefficients. However, due to the potentially large range of wave numbers, the raw \( k \)-values are not fed directly into the network. Instead, a logarithmic transformation and scaling are applied to ensure that the input lies in a numerically stable range for training the neural network.

\paragraph{Input Transformation:}  
Given a raw wavenumber \( k \), the input to \texttt{NetNN} is chosen as:
\[
x_{\text{in}} = 0.1 \log(1 + k).
\]
This transformation compresses a wide range of \( k \)-values into a more manageable interval. For small \( k \approx 0.1 \), \(\log(1 + k)\) is close to zero, and for large \( k \), the logarithmic growth ensures that the input does not become excessively large. The factor of \(0.1\) further scales the logarithmic value into a range that neural network layers can handle effectively without saturation or numerical instability.

\paragraph{Nonlinear Output Scaling and Asymptotic Correctness:}  
After \texttt{NetNN} produces its raw outputs, the code applies a nonlinear scaling,
\begin{equation}\label{nonlinscaling}
    N_{\text{out}} = \texttt{Net2}(0.1\log(1 + k)) \cdot \frac{2\,\exp\bigl(-2/(k+10^{-2})\bigr)}{1+\exp\bigl(-2/(k+10^{-2})\bigr)},
\end{equation}
where the logistic-type factor in \eqref{nonlinscaling} smoothly modulates the amplitude of the learned corrections as a function of \( k \). For small \( k \), corresponding to the long-wavelength limit, this scaling ensures that the generalized transport coefficients approach the classical Navier--Stokes asymptotics.
For larger \( k \), the neural corrections can deviate more substantially, capturing rarefaction and non-Newtonian effects.

\paragraph{Computation of \(\tau_i\):}  
Given \(N_{\text{out}}\), the next step is to separate the components according to
\[
N_2(k) = N_{\text{out},1}, \quad N_6(k) = N_{\text{out},2}.
\]
Then, the generalized transport coefficients are given by
\begin{equation}
\begin{split}
    \tau_1 & = -k,\quad \tau_2 = -\frac{4}{3}\mathrm{Kn}\, k^2 \exp(N_2(k)),\\
\tau_3 & = -k,\quad \tau_4 = 0,\quad \tau_5 = -\frac{2}{3}k,\\
\tau_6 & = -\frac{5}{3}\mathrm{Kn}\, k^2 \exp(N_2(k)+N_6(k)). 
\end{split}  
\end{equation}


These relationships define, after appropriate scaling and transformations, the output of the neural network, thus yielding the generalized transport coefficients \(\tau_i\) that reflect modified hydrodynamic behavior in the rarefied regime.

\subsection*{Determination of Training Weights}

In the training procedure, we minimize a weighted mean-squared error between predicted and reference spectra. For each wave number \( k \). Denote
\[
M(k) = \max_{\omega} \rho_{\text{spec}}(k,\omega),
\]
where \(\rho_{\text{spec}}(k,\omega)\) is the reference density fluctuation spectrum at wave number \( k \) and frequency \( \omega \). The weight \( W(k,\omega) \) is chosen as
\[
W(k,\omega) = \frac{1}{M(k)^2 [\,1 + 0.2\,k\,]},
\]
which ensures that:
\begin{enumerate}
    \item Larger spectral magnitudes do not disproportionately dominate the training loss, as we normalize by \( M(k)^2 \).
    \item The factor \( 1/(1+0.2k) \) avoids overemphasis on large \( k \)-values, leading to a more balanced training across scales.
\end{enumerate}


\section{Computing Time Evolution from Frequency--Wave number Spectra}
\label{AppTimeEvolution}

In this appendix, we describe how to compute the time evolution of density and temperature fields from a particular initial density profile. The presented method relies on interpreting the density fluctuation spectra as Green’s functions. Indeed, the spatio-temporal Fourier transform of the density \(\tilde{\rho}(\omega,k)\) as obtained from the neural network is the unique solution to the density evolution with initial condition \(\rho_0(\mathbf{x}) = \delta(\mathbf{x})\), where $\delta$ is Dirac's delta distribution, see also Appendix \ref{AppLightScattering}. Thus, \(\langle \rho^2 \rangle(\omega,\mathbf{k})\) can be interpreted as the Green’s function \(G(\omega,\mathbf{k})\) describing the response of the system to a unit impulse at the initial time.

Let
\[
\hat{\rho}_0(\mathbf{k}) = \frac{1}{(2\pi)^{3/2}} \int_{\mathbb{R}^3} \rho_0(\mathbf{x}) e^{-i \mathbf{k} \cdot \mathbf{x}} \, d^3x,
\]
denote the spatial Fourier transform of the initial density profile $\rho_0$.
By linearity, the solution with initial condition $\rho_0$ is given by
\[
\widetilde{\rho}(\omega,\mathbf{k}) = G(\omega,\mathbf{k})\,\hat{\rho}_0(\mathbf{k}),
\]
as a frequency-wave-number-representation of the density solution. Once \(\widetilde{\rho}(\omega,\mathbf{k})\) is known, we can invert the frequency transform to obtain \(\rho(t,\mathbf{k})\) in physical variables, 
\[
\hat{\rho}(t,\mathbf{k}) = \frac{1}{\sqrt{2\pi}} \int_{0}^{\infty} \widetilde{\rho}(\omega,\mathbf{k}) e^{i \omega t} \, d\omega,
\]
using a one-sided inverse Fourier transform. After recovering physical time in \(\hat{\rho}(t,\mathbf{k})\), we recover the full physical density \(\rho(t,\mathbf{x})\) depending on the problems geometry:

\begin{itemize}
    \item For planar or Cartesian problems, a standard inverse Fourier transform gives
\[
\rho(t, \mathbf{x}) = \frac{1}{(2\pi)^{3/2}} \int_{\mathbb{R}^3} \rho(t, \mathbf{k}) e^{i \mathbf{k} \cdot \mathbf{x}} \, d^3k.
\]
    \item For problems in radial coordinates, a spherical Bessel transform is used. If \(\rho(t,k)\) represents a radial spectrum, then
    \[
\rho(t, r)= \frac{4\pi}{(2\pi)^{3/2}}\int_0^{\infty} \rho(t, k) \, k^2 \frac{\sin(k r)}{k r} \, dk.
    \]
Numerically, this can be accomplished using specialized transform routines. 
\end{itemize}

In this work, a spherical Fourier transform package (`pyNumSBT`) is applied after the time-domain signal \(\rho(t,k)\) is obtained, thus yielding \(\rho(t,r)\). Other quantities like temperature field is computed similarly using the density-temperature spectra \(\langle \rho T \rangle(\omega,\mathbf{k})\) solved from \eqref{solve spectra}.

\begin{figure}[h!]
    \centering
    \includegraphics[width=1\linewidth]{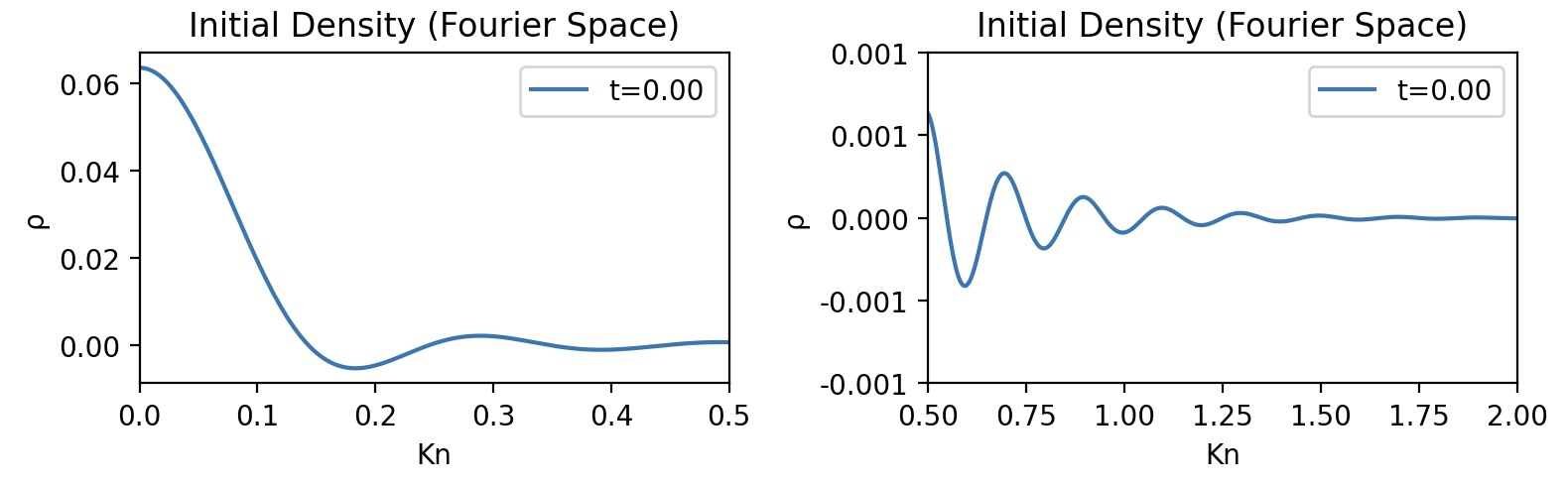}
    \caption{ 
The spatial Fourier transform of the initial density drop, containing frequencies beyond the critical Knudsen number.}
    \label{Lightscatterinsetup}
\end{figure}

\end{document}